%% file: main.tex
\documentclass[final,5p,times,twocolumn]{elsarticle}

\usepackage{amssymb}
\usepackage{lipsum}  
\usepackage{placeins}
\usepackage{subcaption}
\usepackage{overpic}
\usepackage{url}
\graphicspath{{Figure_Updated_simulation/}}
\usepackage[pdftex, pdftitle={Article}, pdfauthor={Author},colorlinks = true,allcolors = blue]{hyperref}
\usepackage{lineno}

\biboptions{numbers,sort&compress}
\begin{document}

\newcommand*{\UCR}{Department of Physics and Astronomy, University of California, Riverside, CA 92521, USA}
\newcommand*{\UCRindex}{1}
\newcommand*{\UCLA}{Department of Physics and Astronomy, University of California, Los Angeles, CA 90095, USA}
\newcommand*{\UCLAindex}{2}

\author[toUCR]{Miguel Arratia\corref{correspondingauthor}}
\cortext[correspondingauthor]{Corresponding author}
\ead{miguela@ucr.edu}
\author[toUCR]{Kenneth Barish}
\author[toUCR]{Liam Blanchard}
\author[toUCLA]{Huan Z. Huang}
\author[toUCLA]{Zhongling Ji}
\author[toUCR]{Bishnu Karki}
\author[toUCR]{Owen Long}
\author[toUCR,toUCLA]{Ryan Milton}
\author[toUCR]{Ananya Paul}
\author[toUCR]{Sebouh J.~Paul}
\author[toUCR]{Sean Preins}
\author[toUCR]{Barak Schmookler}
\author[toUCLA]{Oleg Tsai}
\author[toUCLA]{Zhiwan Xu}

\address[toUCR]{\UCR} 
\address[toUCLA]{\UCLA}

\title{A high-granularity calorimeter insert based on SiPM-on-tile technology\\
at the future Electron-Ion Collider}

%\begin{graphicalabstract}
%\centering
%\includegraphics[width=\textwidth]{endcap_and_insert.png}\\\includegraphics[width=0.55\textwidth]{cross_section.png}\includegraphics[width=0.45\textwidth]{explode_view.png}
%\end{graphicalabstract}
%\date{\today} 

\begin{abstract}
We present a design for a high-granularity calorimeter insert for future experiments at the Electron-Ion Collider (EIC). The sampling-calorimeter design uses scintillator tiles read out with silicon photomultipliers. It maximizes coverage close to the beampipe, while solving challenges arising from the beam-crossing angle and mechanical integration. It yields a compensated response that is linear over the energy range of interest for the EIC. Its energy resolution meets the requirements set in the EIC Yellow Report even with a basic reconstruction algorithm. Moreover, this detector will provide 5D shower data (position, energy, and time), which can be exploited with machine-learning techniques. This detector concept has the potential to unleash the power of imaging calorimetry at the EIC to enable measurements at extreme kinematics in electron-proton and electron-nucleus collisions. 
\end{abstract}

\maketitle
\newpage
\tableofcontents
%\newpage
\input{Introduction}

\input{Design}

\FloatBarrier
\input{Simulations}
%\FloatBarrier
\input{Conclusions}

\newpage

\section*{Model availability}
The Google \textsc{Sketchup} model, as well as the CAD file in .step format, for the proposed detector can be found in Ref.~\cite{sketchupModel}.
The \textsc{DD4HEP} model can be found in Ref.~\cite{Milton_EIC_pEndcap_Insert_2022}.

\section*{Acknowledgements} \label{sec:acknowledgements}
We thank members of the EIC Detector-1 proto-collaboration for useful discussions about the insert, as well as help from their Software Team on matters related to the \textsc{DD4Hep} framework. We thank Oskar Hartbrich for pointing out the importance of the timing cuts and the choice of resolution metric to reproduce CALICE results. We also thank Gerard Visser for feedback on the readout scheme. We also thank members of the California EIC consortium for their feedback on our design. This work was supported by MRPI program of the University of California Office of the President, award number 00010100. M.A, A.P and S.P acknowledge support from DOE grant award number DE-SC0022324.  M.A, K.B, and B.K acknowledge support from DOE grant award number DE-SC0022355. 

\FloatBarrier
\renewcommand\refname{Bibliography}
\bibliographystyle{elsarticle-num}
\bibliography{bibio.bib} % refers to example.bib

\appendix
\end{document}

%% file: Introduction.tex
\section{Introduction}
\label{sec:outline}
One of the key requirements for detectors at the future Electron-Ion Collider (EIC) is to have tracking and full calorimetry with 2$\pi$ azimuthal coverage over a large range in pseudorapidity---nominally $|\eta|<4.0$~\cite{AbdulKhalek:2021gbh}. This would ensure a ``$4\pi$, general-purpose detector'' that could deliver the original EIC scientific goals~\cite{Accardi:2012qut} and much beyond. 

While both the ATHENA~\cite{ATHENA} and ECCE~\cite{ecce_consortium_2022_6537588,Adkins:2022jfp} detector designs contemplate close-to-full coverage with full calorimetry, the specifics on how to implement it in the forward region remain undefined. Covering the region $3<|\eta|<4$ is challenging due to the EIC beam-crossing angle, which is 25~mrad~\cite{AbdulKhalek:2021gbh}. 

\begin{figure*}[h]
    \centering
    \includegraphics[width=0.375\textwidth]{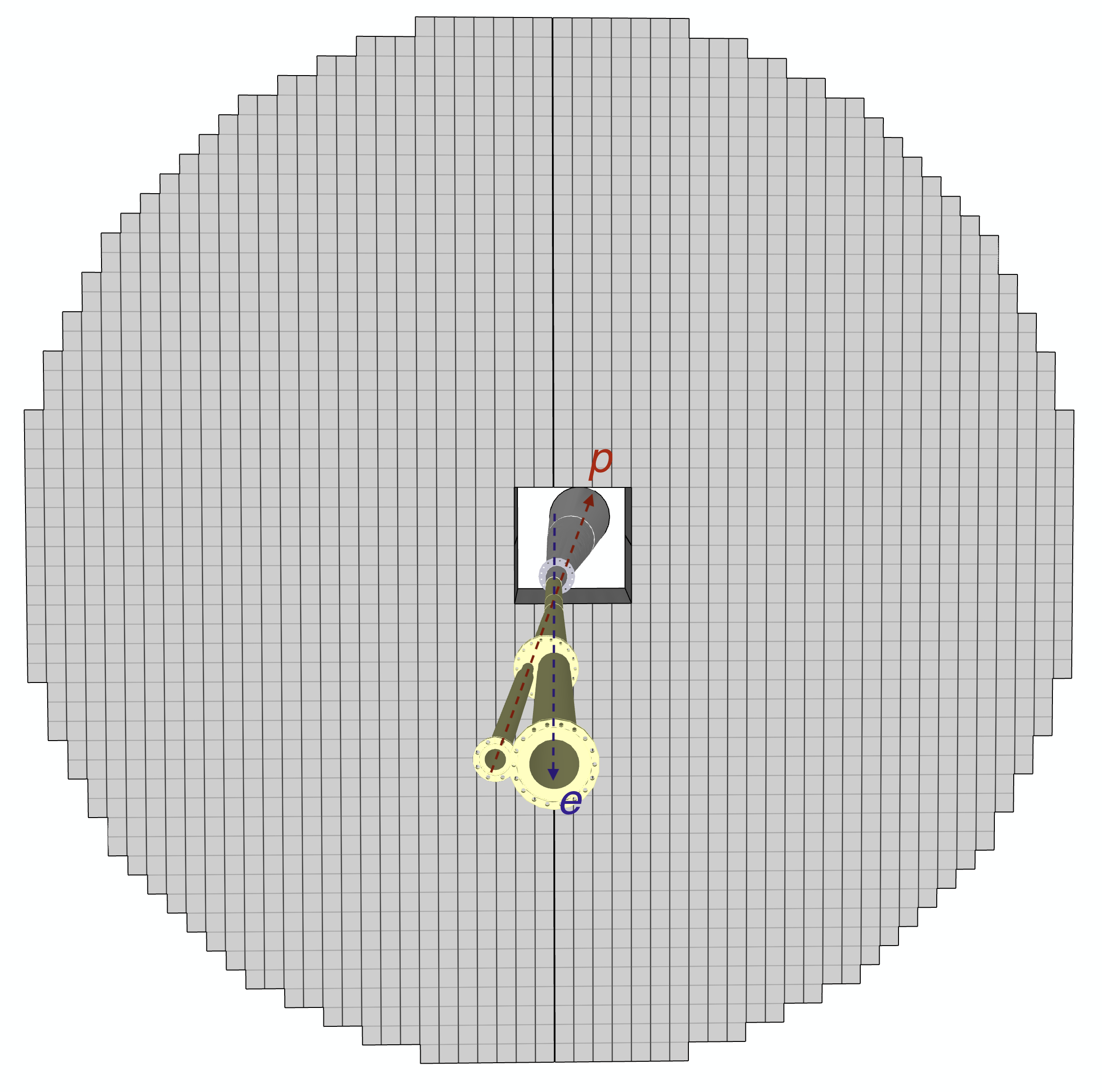}\includegraphics[width=0.3\textwidth]{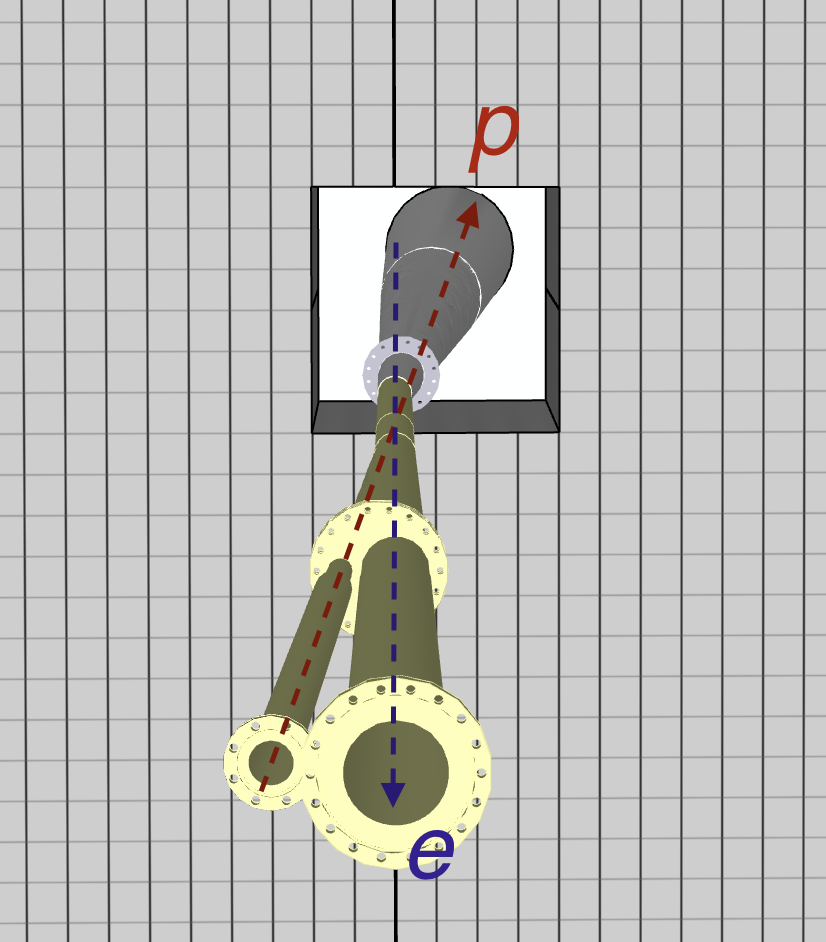}
    \caption{Left: Forward calorimeter endcap and beampipe without the insert, with arrows detailing the directions of the electron and proton beams.  Right:  Same, zoomed in to show details. The beampipe axis is slightly to the left of the proton direction.}
    \label{fig:challenge}
\end{figure*}

As illustrated in Fig.~\ref{fig:challenge}, the beampipe envelope crosses the region where the forward hadronic calorimeter (HCal) would be located at an angle that approximately corresponds to the proton direction. Any detector covering this region needs to simultaneously fill a complex volume and keep clearance from the beampipe, while fitting the other calorimeters without needing additional support structures.

In addition, the detector in the $3<\eta<4$ region needs to be well matched to the particle densities and energies expected at the EIC. At the highest energy setting of {18 GeV} electron beam and {275 GeV} proton beam, the jet spectra reach close to the proton-beam energy with a high rate at nominal luminosity~\cite{AbdulKhalek:2021gbh}. Furthermore, single-particle measurements up to 60 GeV are set as a requirement~\cite{AbdulKhalek:2021gbh}. Given the large energies involved, an HCal becomes crucial since poor tracking performance is expected at forward angles~\cite{Arrington:2021yeb,AbdulKhalek:2021gbh}.

Instrumenting the $3<\eta<4$ region demands a relatively small detector area ({$\approx0.6\times0.6$ m$^{2}$} for a detector located at $z=3.8$ m) but also high granularity to yield a reasonable $\eta$ resolution and disentangle nearby particles. Small cells can also tessellate the complex geometry near the beampipe in an effective manner.

Good angular coverage and resolution would be beneficial for measurements of jets and their substructure, which will enable a wide variety of studies at the EIC, and inclusive deep-inelastic scattering at high-$x$~\cite{AbdulKhalek:2021gbh,AbdulKhalek:2022hcn}. 

The concept of high-granularity calorimetry has been developed for the last two decades or so~\cite{Sefkow:2015hna}, driven by the CALICE collaboration, and motivated by the particle-flow paradigm~\cite{Thomson:2009rp}. More recently, this approach has been adopted for the CMS High-Granularity Calorimeter (HGCAL) upgrade~\cite{CERN-LHCC-2017-023}, and is envisioned for experiments at several proposed $e^{+}e^{-}$ colliders such as ILC~\cite{ILDConceptGroup:2020sfq}, CLIC~\cite{Dannheim:2019rcr}, and CEPC~\cite{CEPCStudyGroup:2018ghi,Li:2021gla,Duan:2021mvk}. 

Extensive R$\&$D efforts~\cite{Sefkow:2015hna} and rapid progress in silicon-photomultiplier (SiPM) technology~\cite{Simon:2018xzl} have culminated in the ``SiPM-on-tile'' concept for hadronic calorimetry. In this approach, the active elements of the calorimeter are small scintillator tiles ($\approx$ 3$\times$ 3 cm$^{2}$) air coupled to a SiPM at their center~\cite{Blazey:2009zz,Simon:2010hf}, without using wavelength-shifting fibers that were used in earlier designs~\cite{Andreev:2004uy,Andreev:2005cua,CALICE:2010fpb,Simon_2010}. 

The advent of high-granularity calorimeters has boosted studies of machine-learning algorithms~\cite{Feickert:2021ajf} that could exploit the benefits of ``imaging calorimetry'' for a variety of tasks~\cite{Paganini:2017dwg,Paganini:2017hrr,Belayneh:2019vyx,Qasim:2019otl,Belayneh:2019vyx,DiBello:2020bas,Buhmann:2020pmy,Akchurin:2021afn,Pata:2021oez,Neubuser:2021uui,Akchurin:2021ahx,Buhmann:2021caf,Khattak:2021ndw,Chadeeva:2022kay,Qasim:2022rww,Mikuni:2022xry}.

Some early proposals for EIC detectors, such as TOPSIDE, considered imaging calorimetry over the entire solid angle~\cite{Repond:2019fbz,AbdulKhalek:2021gbh}, including hadronic calorimetry similar to the CALICE design, although those options were not further pursued\footnote{In particular, they were not in any of the official EIC proposals~\cite{ATHENA,ecce_consortium_2022_6537588,core_proto_collaboration_2021_6536630}. On the other hand, an imaging electromagnetic calorimeter based on silicon-pixel sensors was included in the ATHENA proposal~\cite{ATHENA}. Imaging calorimetry with silicon sensors is also envisioned for the zero-degree calorimeters.}. Instead, the forward HCals for both the ATHENA~\cite{ATHENA} and ECCE~\cite{Bock:2022lwp} proposals included a more traditional steel-scintillator design with longitudinal granularity that varies between 4 and 7 segments, which is similar to that used in the ZEUS~\cite{ZEUSCalorimeterGroup:1989ill} and H1~\cite{H1CalorimeterGroup:1993boq} detectors at HERA.

The SiPM-on-tile technology represents an attractive option to instrument the {$3<\eta<4$} region at the EIC for various reasons. First, the expected energy range for jets largely overlaps with that of proposed $e^{+}e^{-}$ Higgs factories, so the vast CALICE literature and test-beam data~\cite{Andreev:2004uy,Andreev:2005cua,CALICE:2010fpb,CALICE:2010thx,CALICE:2011brp,CALICE:2012eac,Simon:2013zya,CALICE:2013osg,CALICE:2013dkh,CALICE:2014tgv,CALICE:2014xjq,CALICE:2015fpv,CALICE:2016nds,CALICE:2018ibt,Sefkow:2018rhp} could inform designs and validate simulations. Second, the radiation level expected at the EIC is orders of magnitude smaller than the LHC so constraints from SiPM radiation damage are not a limiting factor. Third, the forward geometry limits the detector volume, numbers of channels, and cost, while still covering a significant range in $\eta$--$\phi$ space. 

In this paper, we present a concept for a high-granularity calorimeter insert (HG-CALI) based on the SiPM-on-tile technology that achieves excellent acceptance near the beampipe, while satisfying the mechanical constraints imposed by integration with the rest of the detector. It is designed to be compensated ($e/h\approx1$) and meets the requirements set in the EIC Yellow Report~\cite{AbdulKhalek:2021gbh} with a basic reconstruction algorithm. Moreover, it will open up potential for improvements based on machine-learning techniques that exploit the 5D measurement of showers. It would be complemented with a high-granularity electromagnetic calorimeter insert based on tungsten powder/scintillating fiber technology~\cite{Tsai_2012}, which will be described in a separate publication.

%% file: Design.tex
\section{Proposed Design}
\label{sec:design}
\subsection{External constraints}
\begin{figure*}[h]
    \centering
    \includegraphics[width=.92\textwidth]{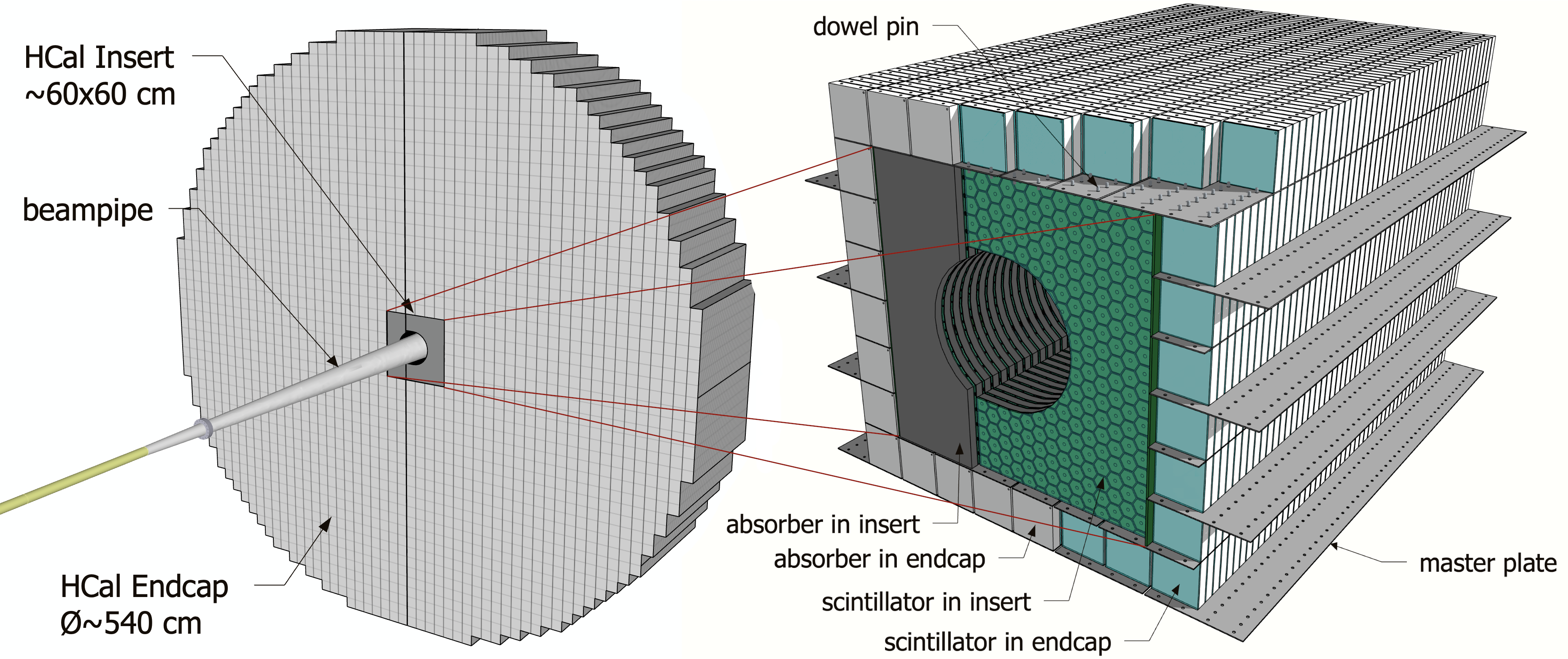}
    \caption{Left: the proposed HCal insert inside the full HCal endcap (as viewed from upstream). Right: Zoom-in of the insert surrounded by only the endcap towers that are adjacent to it. The beampipe, and some of the absorbers and dowel pins are hidden to show how the insert connects to the rest of the detector. }
    \label{fig:endcap}
\end{figure*}

Two main factors constrain the design of the proposed calorimeter insert. First, it must integrate into the larger calorimeter endcap without additional support structures that would create dead areas. Second, it must surround the beampipe within some minimal acceptable clearance in order to maximize its coverage. 

In the ATHENA design~\cite{ATHENA}, the forward HCal design is based on the technology previously used in the STAR experiment~\cite{Tsai:2015bna,Aschenauer:2016our} and consists of alternating layers of steel absorbers (consisting of $96\times98\times 20$~mm blocks) and scintillator tiles with a thickness of 3~mm. The absorber blocks of both the endcap and the insert are held in place using dowel pins which go into the holes on the top of each absorber block, through holes of the 1.9~mm thick ``master plates'', and then into the holes in the blocks above them. This is illustrated in Fig.~\ref{fig:endcap}, which shows the high-granularity insert (described in more detail in the following sections) surrounded by the full HCal endcap (left), and zoomed-in with some of the adjacent structures within the endcap (right). 

The endcap HCal is designed to open up by splitting in half vertically for maintenance, as it will rest on a support structure with a rail system~\cite{AbdulKhalek:2021gbh}. Therefore, the insert is designed to split into two parts along with the endcap. 

The current design for the beampipe~\cite{beampipe} is conically shaped and has an outer radius of 11.15~cm at the upstream face of the HCal ($z=380$~cm) and 13.40~cm at the front face of the last layer of the HCal ($z=498$~cm). Furthermore, the beampipe axis is at a small angle ($-$24.3~mrad) relative to the lab-frame $z$ axis, which is defined parallel to the solenoidal magnetic field axis and to the electron beampipe. The proton beam axis is at $-$25~mrad relative to the electron axis. Distances and angles defined with respect to the proton axis are labelled with an asterisk superscript throughout this paper. We assume a clearance of about {4~cm} along the entire volume of the HCal insert. 

Based on these requirements, the proposed HG-CALI will cover the equivalent area of 6$\times$6 blocks of the HCal endcap, or approximately 60$\times$60~cm. However, since the beampipe is offset in the negative $x$ direction, the part of the insert on the left side when viewed from upstream (positive $x$) covers the width of two HCal endcap blocks, plus the gaps between them, (19.6~cm), while the right side (negative $x$) of the insert covers the width of four HCal endcap blocks and the gaps between them (39.6~cm). Together with the 4~mm gap between the two halves of the HCal, and the 4~mm of readout backplanes on either side, the total width of the proposed insert is 60.4~cm. 

\begin{figure}[h]
    \centering
    \includegraphics[width=\columnwidth]{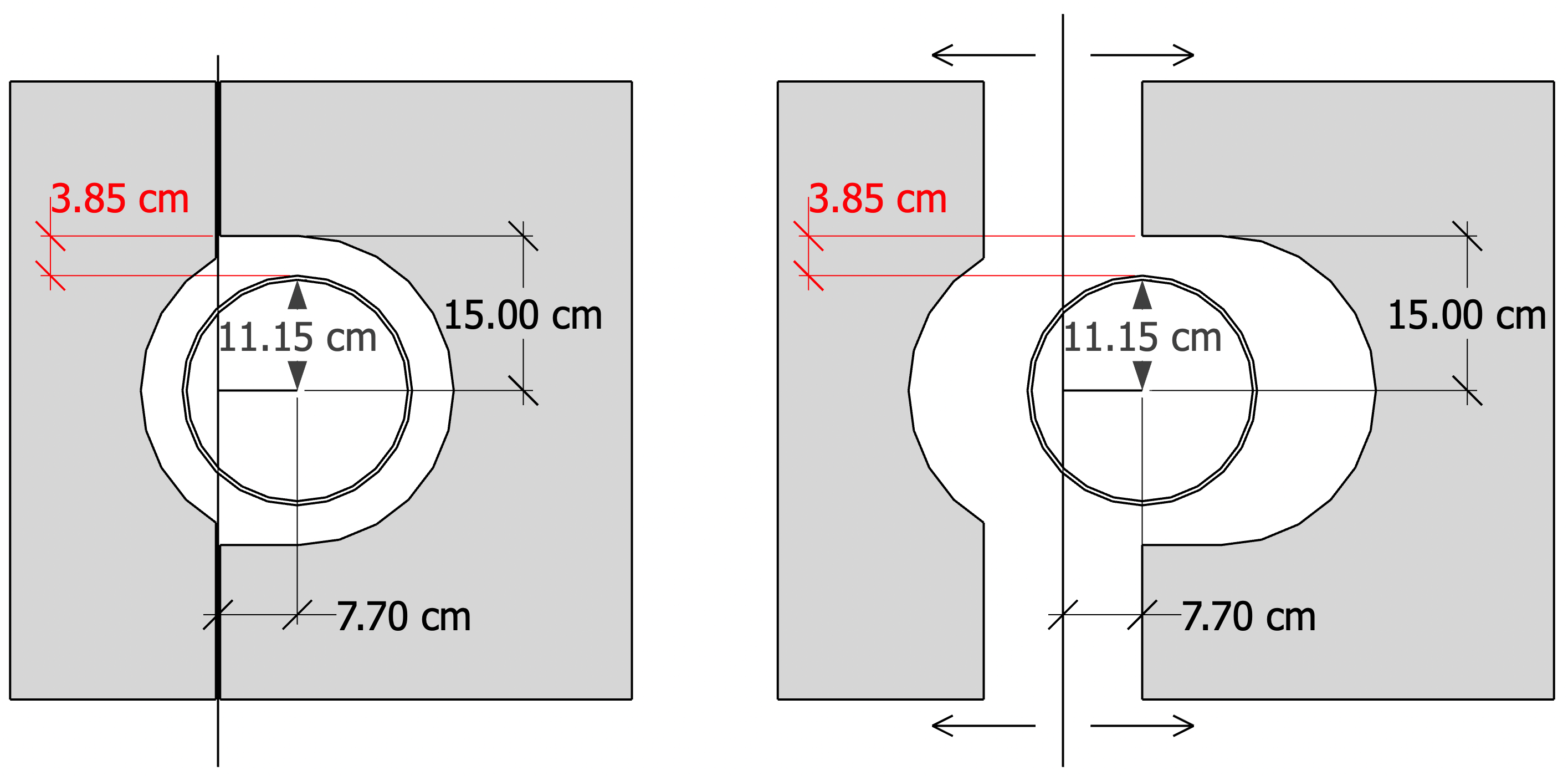}
    \caption{Transverse cross section of the first layer of the insert, detailing the shape of the hole surrounding the beampipe. Left: The cross section of the insert when it is in the closed configuration. Right: The insert as it is being opened, illustrating the need for the D shape of the hole on the right part of the detector in order to achieve clearance even while its being opened.}
    \label{fig:clearance_definition}
\end{figure}

To account for the beampipe, we include a hole in each layer of the insert, as shown in the left side of Fig.~\ref{fig:clearance_definition}. 
The shape of this hole is circular for the most part. However, the right side of the insert is shaped like a capital letter D, with flat extensions on the top and bottom to ensure clearance even while the calorimeter is sliding open.

\begin{figure}[h]
    \centering
    \includegraphics[width=\columnwidth]{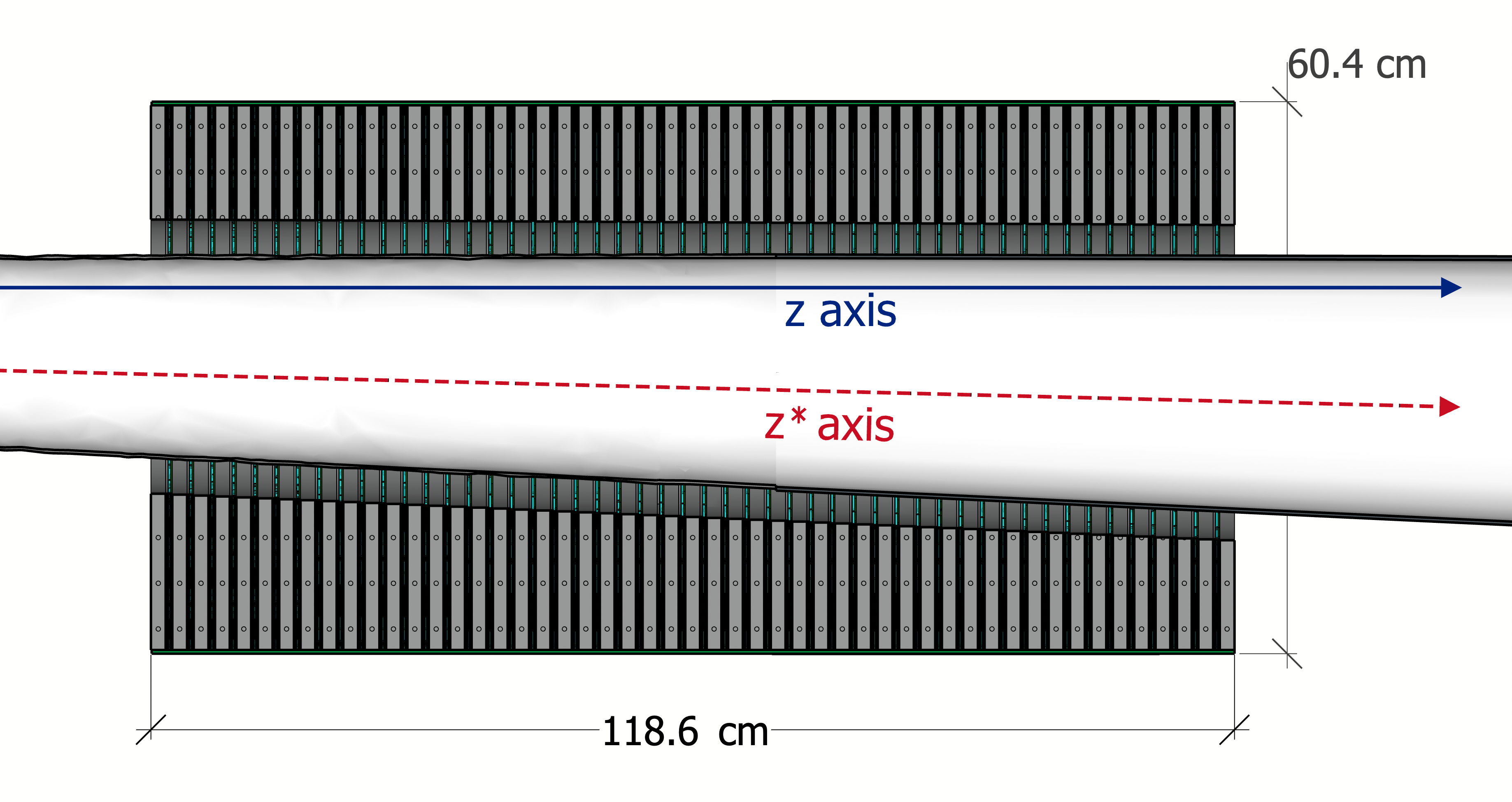}\\
    \hspace{0.03\textwidth}\includegraphics[width=\columnwidth]{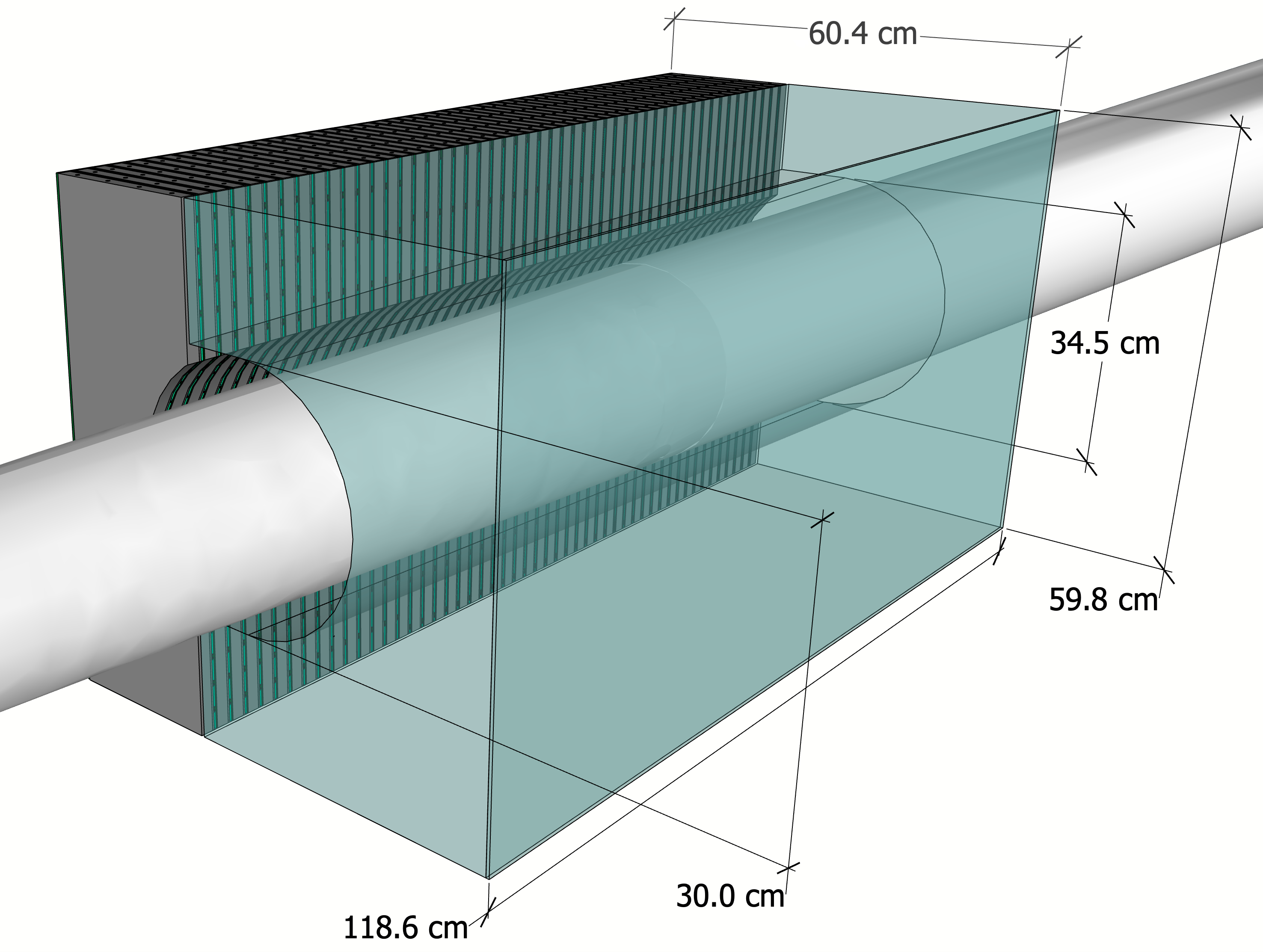}
    \caption{Top: Cross section of the beampipe and the calorimeter insert surrounding it, as viewed from above. The $z$ (anti-parallel to the electron beam) and $z^*$ (proton beam) axes are indicated as a solid line (blue online) and a dashed line (red online), respectively.  Bottom: 3D view of the insert with one side replaced with a wireframe to show the beampipe inside of it.}
    \label{fig:cross_section}
\end{figure}

The radii and position of the centers of the circles vary from layer to layer, as shown in the top image of Fig.~\ref{fig:cross_section}, because of the conical shape and tilt (respectively) of the beampipe. This approach leads to an optimal layout, in the sense that it covers all the available volume, being limited only by the clearance from the beampipe. The dimensions of the insert are shown in the bottom image of Fig.~\ref{fig:cross_section}.

Although we have shown a specific way to implement the HG-CALI with the ATHENA HCal design~\cite{ATHENA}, this could be adapted to fit the ECCE HCal~\cite{Bock:2022lwp} design, which also uses a modular assembly structure based on rectilinear towers that run parallel to the $z$ axis. The HG-CALI could also be incorporated into the second EIC detector~\cite{eicug_2022_6422182}; for example, it is also compatible with the CORE design~\cite{core_proto_collaboration_2021_6536630,CORE:2022rso}. It could also be used in future experiments at the Chinese EicC, which considers a crossing-angle of 50~mrad~\cite{Anderle:2021wcy}.

\subsection{Sampling calorimeter layout}
\begin{figure*}[h]
    \centering
    \includegraphics[width=0.98\columnwidth]{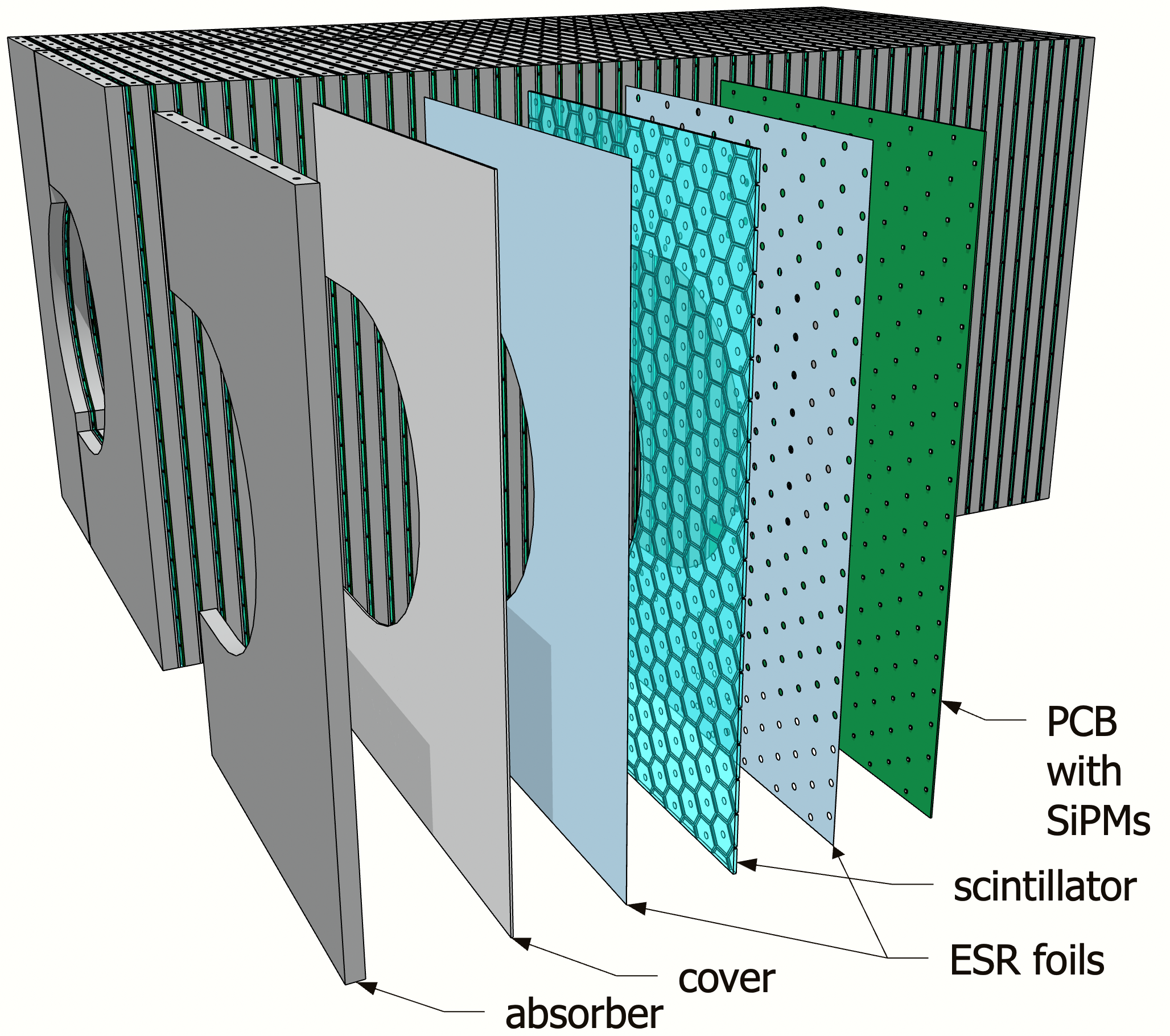}
    \includegraphics[width=\columnwidth,trim={25cm 0 25cm 6cm }, clip]{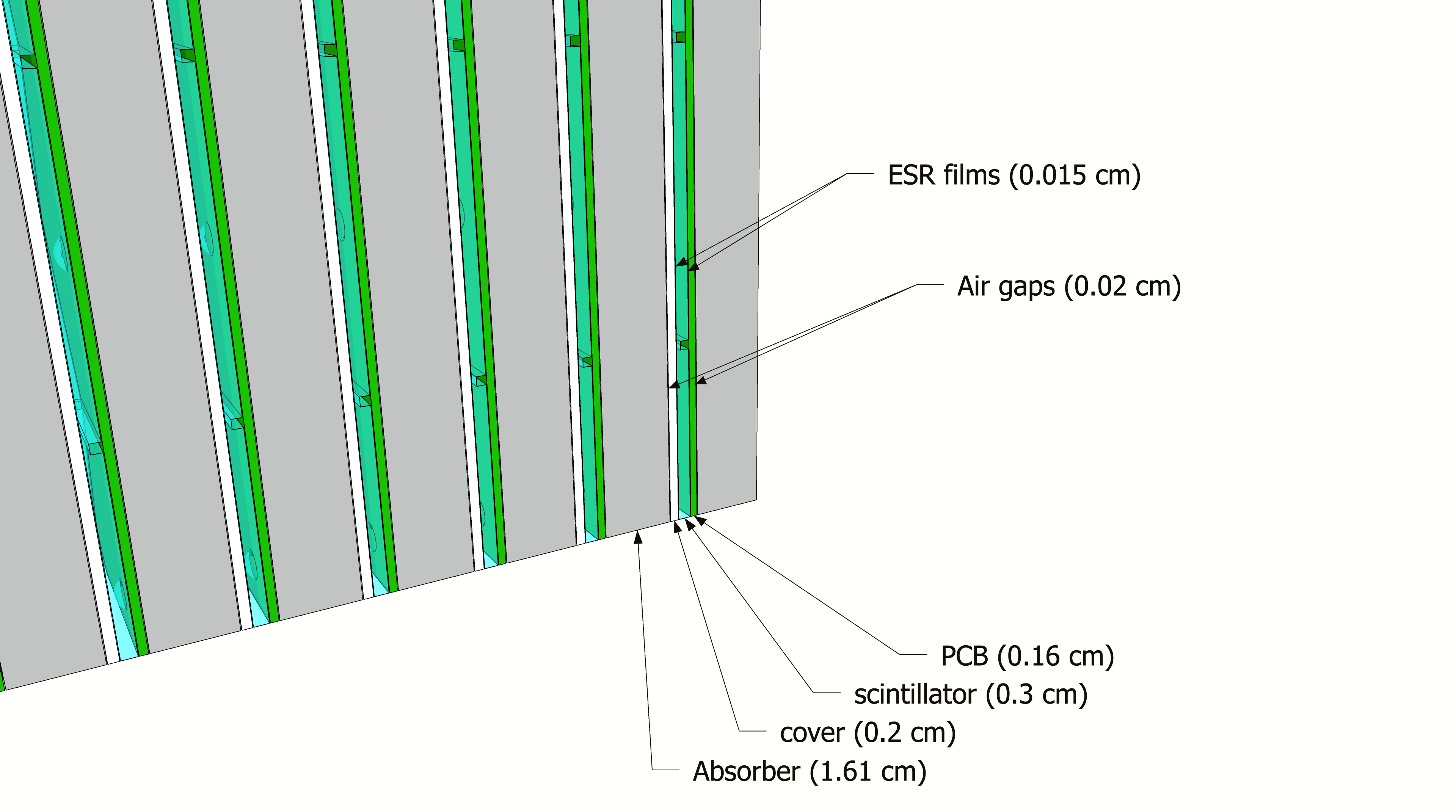}
    \caption{Left: exploded view of the components within a layer of the insert. Right: A zoom-in of one corner of the insert, detailing the longitudinal dimensions of each component of the layers. These include the air gaps (0.02~cm) between the absorber and the scintillator covers and between the absorbers and the PCB and the 0.015~cm ESR foil on either side of the scintillators.  In total, each layer is 2.34~cm thick.}
    \label{fig:explode_view}
\end{figure*}

The HG-CALI consists of alternating absorber blocks and scintillator assemblies. Each scintillator assembly consist of a plastic scintillator sandwiched between a 2~mm cover on the upstream side of it and a printed-circuit board (PCB) with SiPMs for readout on the downstream side. It has 50 layers, each of which is 2.34~cm thick, followed by a downstream absorber-only layer. Figure~\ref{fig:explode_view} shows the longitudinal dimensions of the absorbers and the different components of the scintillator assemblies.

The HG-CALI is divided into two sets of layers with different absorber materials. In the first, tungsten is used; in the second, high-strength steel is used. The use of tungsten is motivated for two reasons: first, it allows for a compensated design; second, its lower radiation length and nuclear-interaction lengths compared to steel ($X_{0}=$ 0.35~cm, $\lambda_{n}$= 9.95~cm for W and $X_{0}=$ 1.76~cm, $\lambda_{n}$= 16.8~cm for Fe) helps in reducing leakage of the core of hadronic showers to the beampipe. 

The use of magnetic steel is required to contain magnetic-field flux. The EIC poses a strict requirement on fringe fields, mainly driven by the need of accurate polarization of the beams. The amount of steel required in this region is still under study and no formal requirement has been established. We have assumed a total thickness of 33.81~cm of steel (the last 21 layers) for the simulations presented in Sec.~\ref{sec:simulation}.

\subsection{Scintillator cells}

A single scintillator ``megatile'' is used for each half-layer, which has a slightly different shape to match the corresponding absorber shape. The scintillator panels are tessellated with hexagonal cells defined by grooves. 

Each cell will have a dimple in its center above the SiPMs that are hosted on the PCB.  The grooves are filled with a mixture of epoxy and titanium dioxide pigment, following procedures similar to the STAR electromagnetic calorimeter~\cite{STAR:2002ymp}.  An alternative could be to paint the cells' edges with reflective paint, such as Saint Gobain BC-621. The latter approach was used in the CMS HCAL~\cite{CERN-LHCC-97-031}. 

To fully tessellate each layer, some of the cells on the edges will be cropped or extended. This is not necessarily an issue given that the design allows for a cell-to-cell calibration based on minimum-ionizing particles. 

The megatile is covered on the front and back with reflective foil like the VM2000 enhanced specular reflector (ESR) to increase the light yield and to improve uniformity across the cell area.  The ESR foil on the downstream side of the megatiles will include holes for the dimples in each cell. The ESR foil will not be wrapped to facilitate assembly; instead, the edges of the scintillator megatiles will be painted. 

\begin{figure}[h]
    \centering
    \includegraphics[width=0.38\columnwidth ]{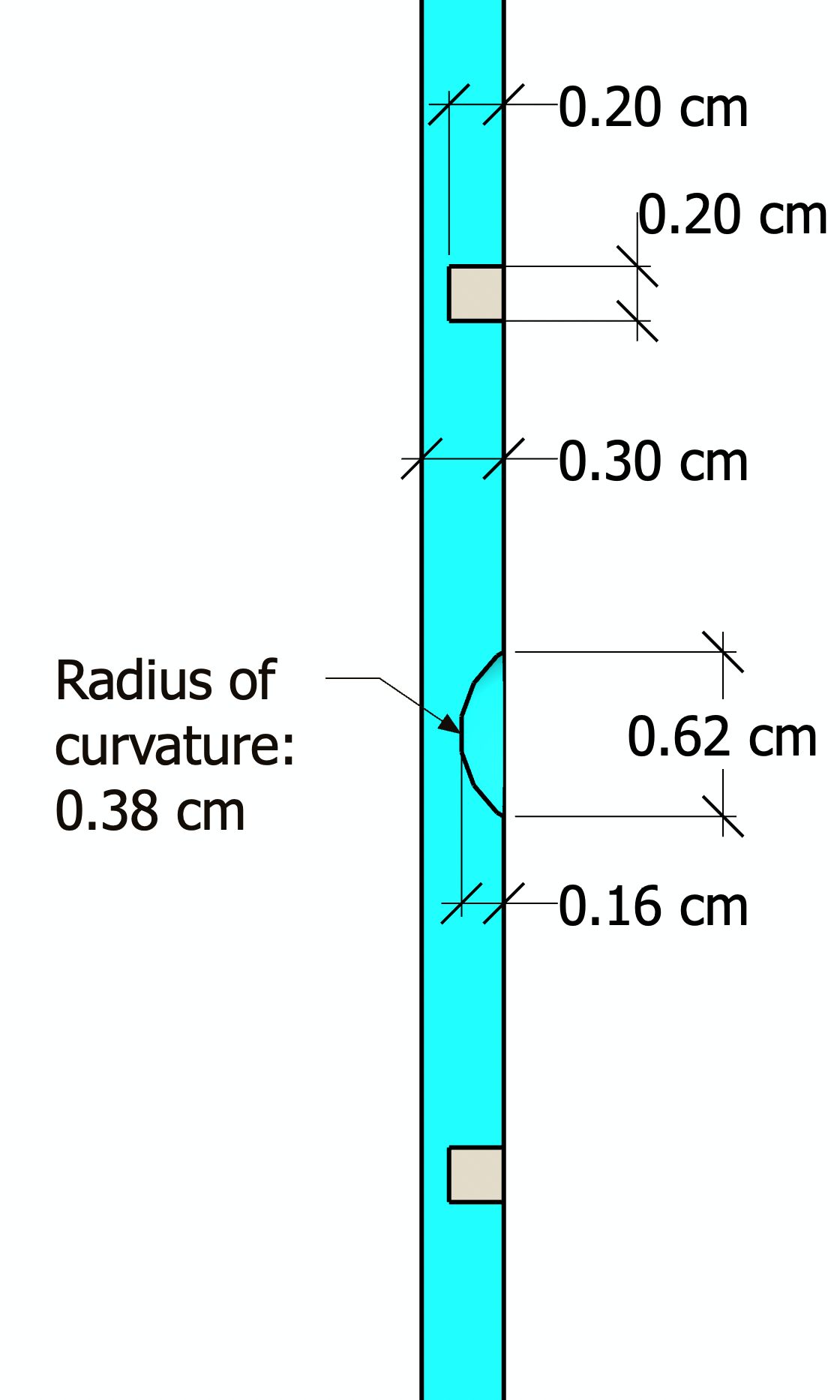}\includegraphics[width=0.62\columnwidth, trim={30cm 0 30cm 0 }, clip]{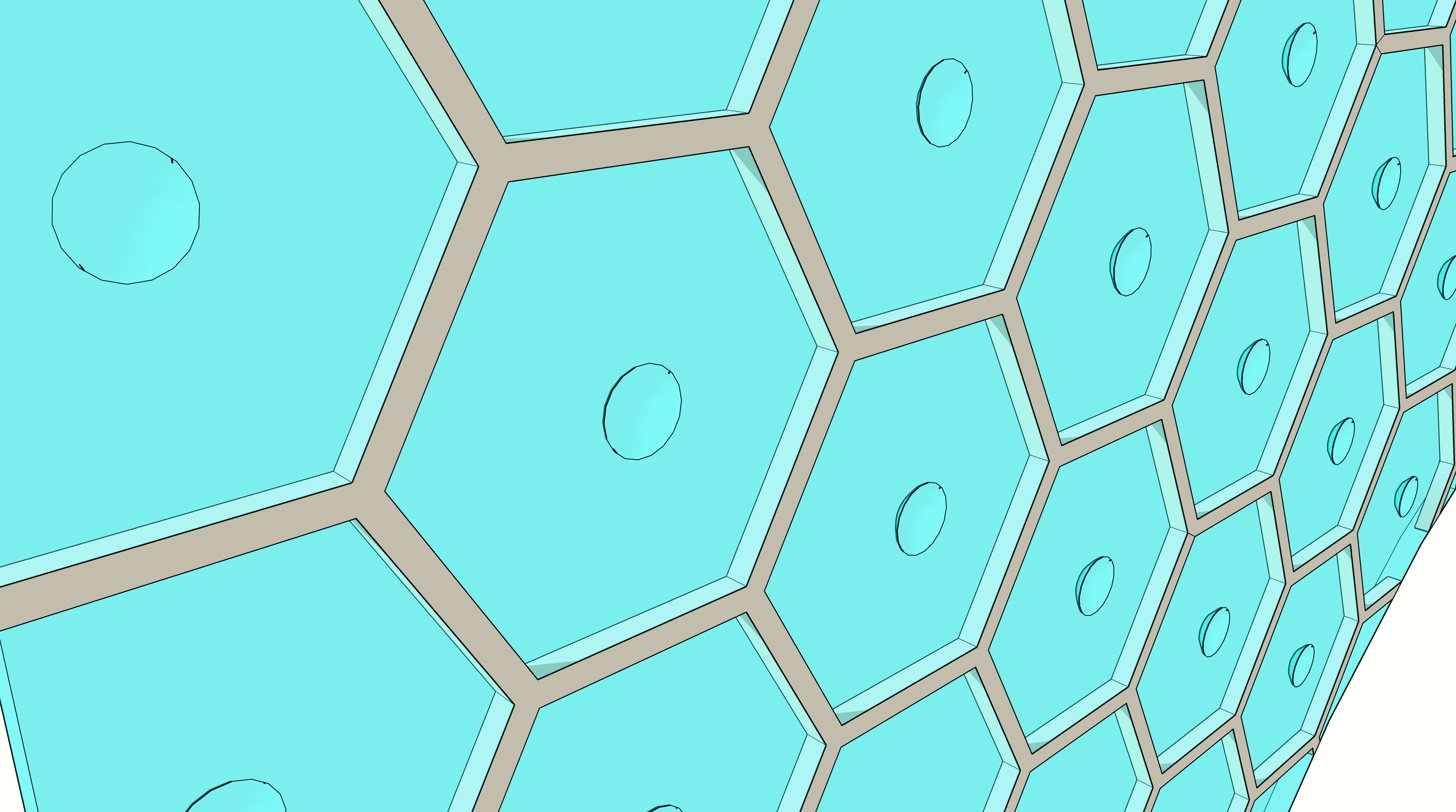}
    \caption{Left:  cross-section view of one scintillator layer, detailing the dimensions of the grooves and dimples thereof. The dimple design follows the dimensions used in Ref.~\cite{Belloni:2021kcw}. Right: Zoom-in of one of the scintillators tiles.}
    \label{fig:grooves_and_dimples}
\end{figure}

The dimples in the center of each cell will have the same dimensions used in the CMS HGCAL test beam~\cite{Belloni:2021kcw}, which define a spherical dome with a radius of curvature of 3.8~mm and a depth of 1.6~mm, as shown in Fig.~\ref{fig:grooves_and_dimples}. The dimples improve signal uniformity across the area of the cell by alleviating hot-spot effects near the SiPM~\cite{Blazey:2009zz,Simon:2010hf}.

The megatile concept has several advantages over a design in which each tile is wrapped individually, such as the ones used in the CALICE prototype~\cite{CALICE:2010fpb} and the CMS HGCAL~\cite{CERN-LHCC-2017-023}, especially for a small and complicated geometry such as that of HG-CALI. The megatile design would simplify assembly and reduce cell-to-cell variations caused by wrapping with ESR foil~\cite{deSilva:2020mak}. 

Previous experience with calorimeters that use megatiles such as the STAR electromagnetic calorimeter~\cite{STAR:2002ymp} suggest that optical cross-talk can be reduced to about 2$\%$ when using a mixture of epoxy and titanium dioxide to fill the grooves that define the cells. This level of optical crosstalk is acceptable for this application, given that hadronic showers encompass a large number of hits over a wide area. 

\begin{figure*}
    \centering
    \includegraphics[width=0.9\textwidth]{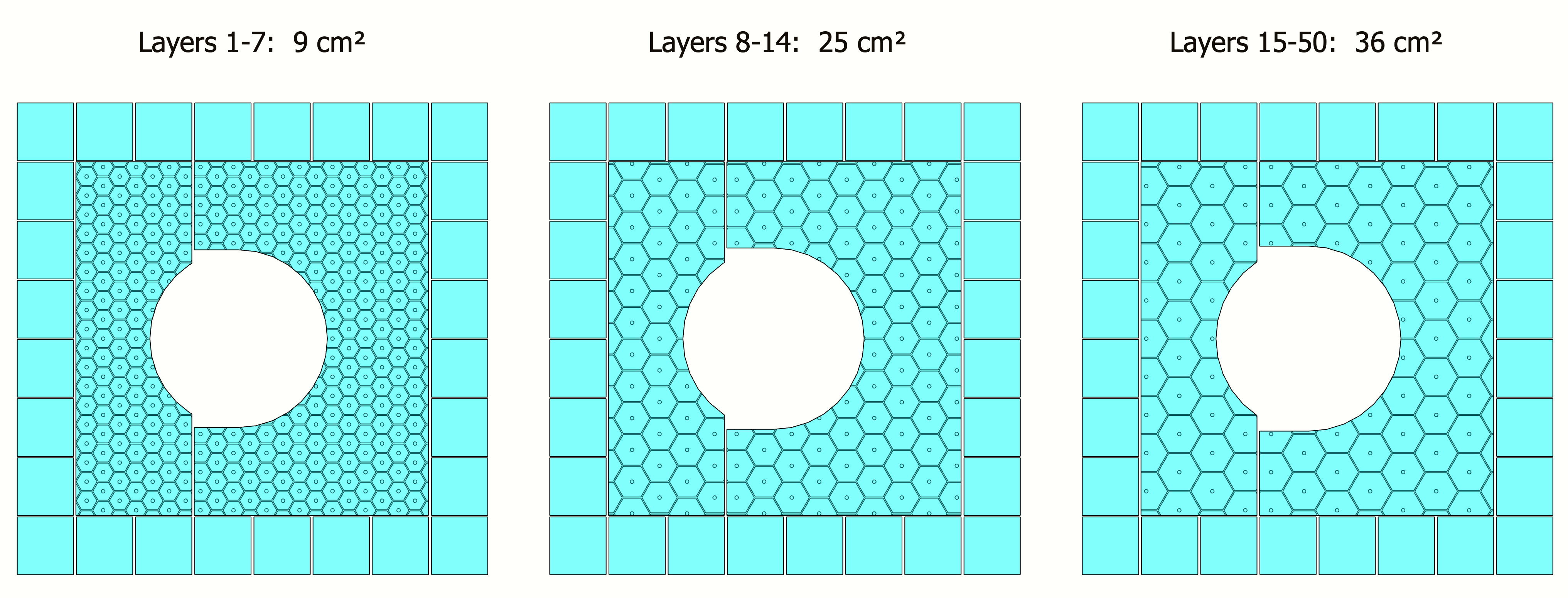}
    \caption{Comparison between the levels of granularity of the insert and the surrounding HCal endcap.  Since the granularity of the insert varies from layer to layer, we show from left to right layers 1, 8, and 15, which respectively represent fine, medium, and coarse granularity.}
    \label{fig:granularity}
\end{figure*}

The megatile concept provides us with the flexibility to define the transverse granularity of the detector, and thus optimize it. We envision that the area of these cells will vary from layer to layer, for example as shown in Fig.~\ref{fig:granularity}, with finer granularity in the furthest upstream layers and coarser granularity in the downstream. The granularity is yet to be optimized, but we estimate that a reasonable choice would yield a total of less than 5k channels.

Test-beam results by the CMS HGCAL collaboration~\cite{Belloni:2021kcw} showed that the light yield of 3~mm thick scintillator cells (Eljen
Technology EJ-200) is about 32~photo-electrons for 9~cm$^{2}$ cells; about 26~photo-electrons for 12~cm$^{2}$ cells, and about 20~photoelectrons for 36~cm$^{2}$ cells. 

\subsection{Silicon photo-multipliers}
Both the CALICE and CMS collaborations use 1.3$\times$1.3~mm$^{2}$ SiPMs~\cite{Belloni:2021kcw,CERN-LHCC-2017-023}. Larger-area SIPMs were considered, \textit{e.g.}, by Jiang \textit{et al.}~in the context of R$\&$D for the CEPC~\cite{Jiang:2020rhv}. Unlike the applications for ILC or CEPC, which will use millions of channels, or the CMS HGCAL, which uses about 200k channels, our design will have just a few thousand channels. Therefore, the total cost of SiPM would still be reasonable even if we choose models with larger areas. 

We anticipate that larger area SiPMs will make the calorimeter more robust against non-uniformity created by SiPM misalignment~\cite{deSilva:2020mak}, as well as saturation effects. The photoelectron yield increases linearly with the area of the SiPM, so choosing a 9~mm$^{2}$ SiPM could increase the light yield by roughly a factor of 4.5 with respect to the CMS HGCAL~\cite{Belloni:2021kcw}. 

As we will show in Sec.~\ref{sec:simulation}, the maximum hit energy reaches $\approx 200$~MeV or 250~MIPs. We anticipate an operation for 60~photo-electrons per MIP for the $3\times3$~cm$^{2}$ cells. This high light yield would create a buffer to cover for the reduced signal-to-noise ratio after radiation damage.

The \textsc{Hamamatsu} model S14160-3025 has about 40k pixels, which translates into a dynamic range of more than 670~MIPs, so no saturation effect is expected for $3\times3$~cm$^{2}$ cells. 

The neutron fluence expected at the EIC near the insert region will not exceed $10^{12}$~neutrons/cm$^{2}$ per year from the electron-proton collisions at the top luminosity $(10^{34}$~cm$^{-2}$s$^{-1} )$~\cite{AbdulKhalek:2021gbh}. The corresponding ionizing radiation dose is 25 Gy per year.

The expected SiPM radiation damage would be classified as ``modest'' in Garutti's review~\cite{Garutti:2018hfu} on the topic. It would mainly lead to increased dark current that can affect the ability to identify single-photon peaks. This level of damage can partially be recovered with annealing~\cite{Garutti:2018hfu}, which could be done after each run (as explained in Sec.~\ref{sec:maintenance}).

For reference, CMS uses SiPM-on-tile technology for regions where the total dose is less than 3000~Gy and fluences less than 8$\times$10$^{13}$~$n_{\mathrm{eq}}$/cm$^{2}$, although the SiPMs will be operated at $-$30 C to reduce the impact of the increased noise~\cite{CERN-LHCC-2017-023}. 

The STAR experiment at RHIC is currently operating a steel-scintillator calorimeter readout with SiPMs~\cite{Tsai:2015bna} that is expected to receive fluences that are similar to those expected at the EIC. Detailed data on radiation-induced noise taken with STAR will guide future iterations of the HG-CALI design. 

\subsection{SiPM-carrying boards and backplane board}
The SiPMs will be hosted in a PCB with no active elements. The PCB will also host LEDs next to each SiPM in the dimple area to be used for monitoring and calibration purposes, similarly to the CMS HGCAL design~\cite{CERN-LHCC-2017-023}. The SiPM-carrying board will host a high-density connector to connect to the backplane board. The SiPMs will not be actively cooled.

\begin{figure}[h]
    \centering
        \includegraphics[width=0.45\textwidth]{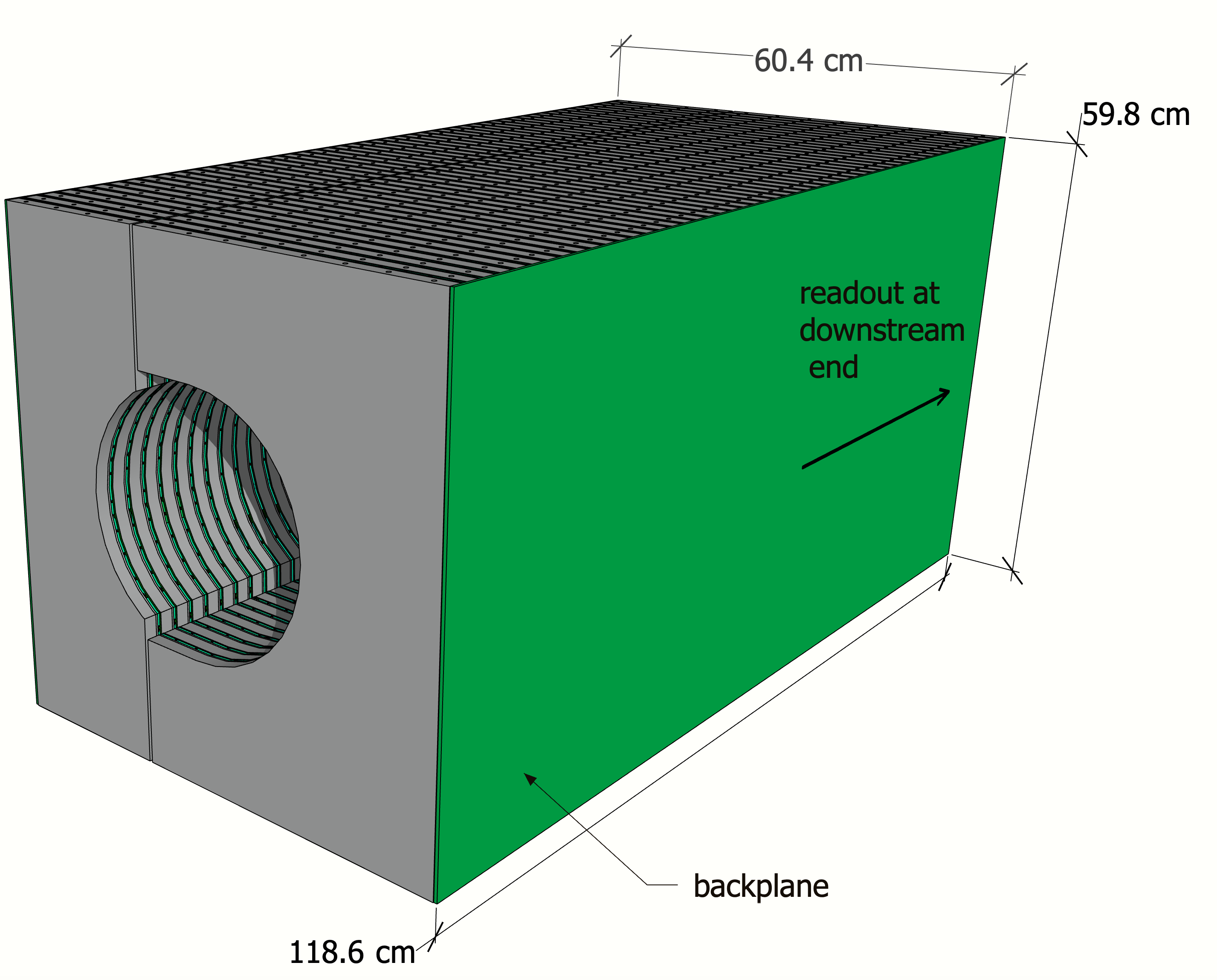}
    \caption{The proposed calorimeter insert, consisting of alternating absorbers and scintillator assemblies, with PCB backplanes on either side of the detector.}
    \label{fig:backplane}
\end{figure}

On either side of the insert are 4~mm PCB backplanes used for readout, which the PCB for each scintillator assemblies connect into like a traditional crate. Figure~\ref{fig:backplane} shows the location of the PCB backplanes. 

The SiPM-carrying boards will be plugged in and unplugged one-at-a-time during assembly and maintenance. The backplane board will also be used to power the SiPMs and control the LED system for calibration and monitoring.

\subsection{Accessibility for maintenance}
\label{sec:maintenance}
\begin{figure}[h!]
     \centering
     \includegraphics[width=0.499\textwidth]{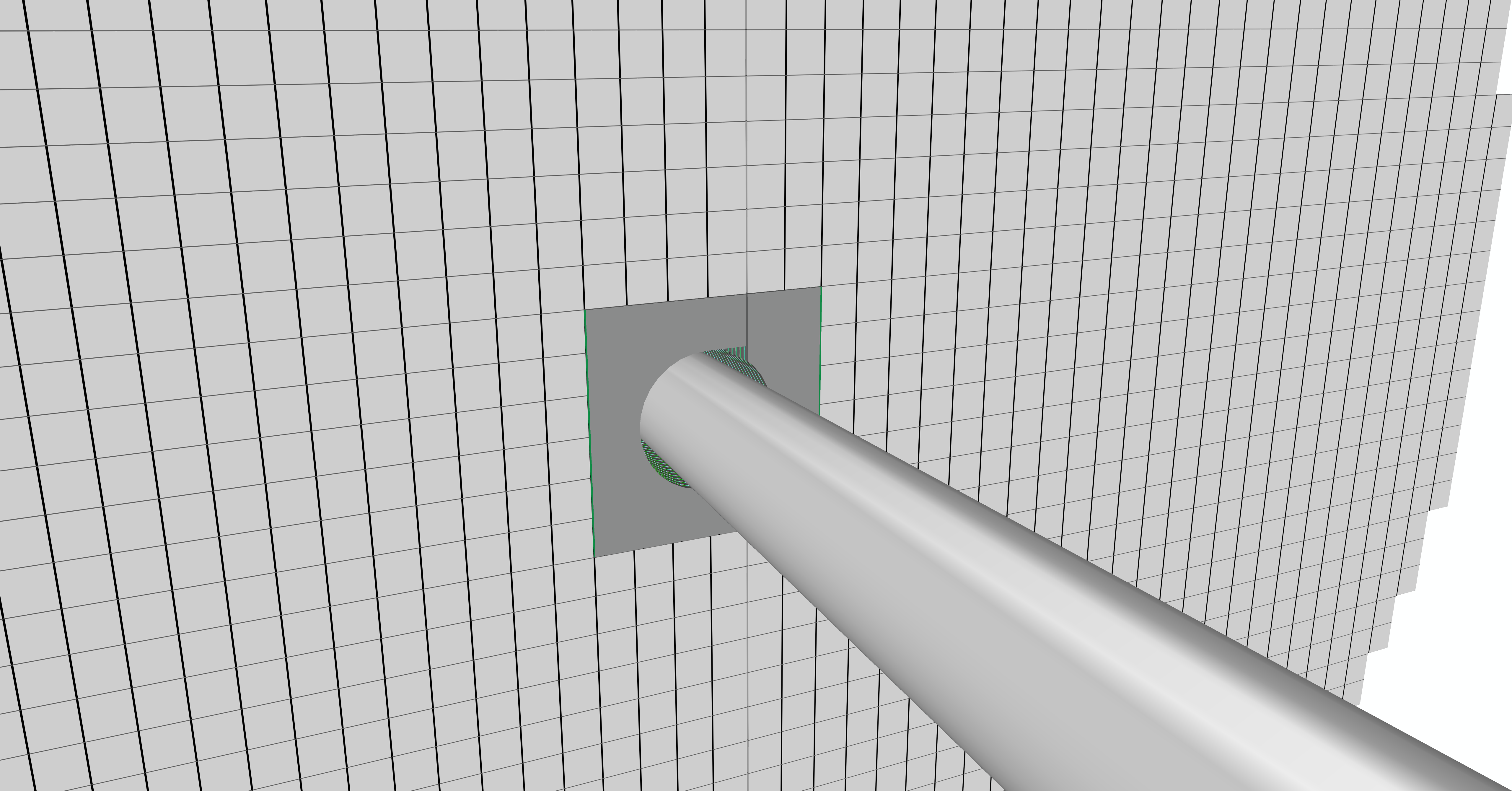}\hspace{.1cm}
     \includegraphics[width=0.499\textwidth]{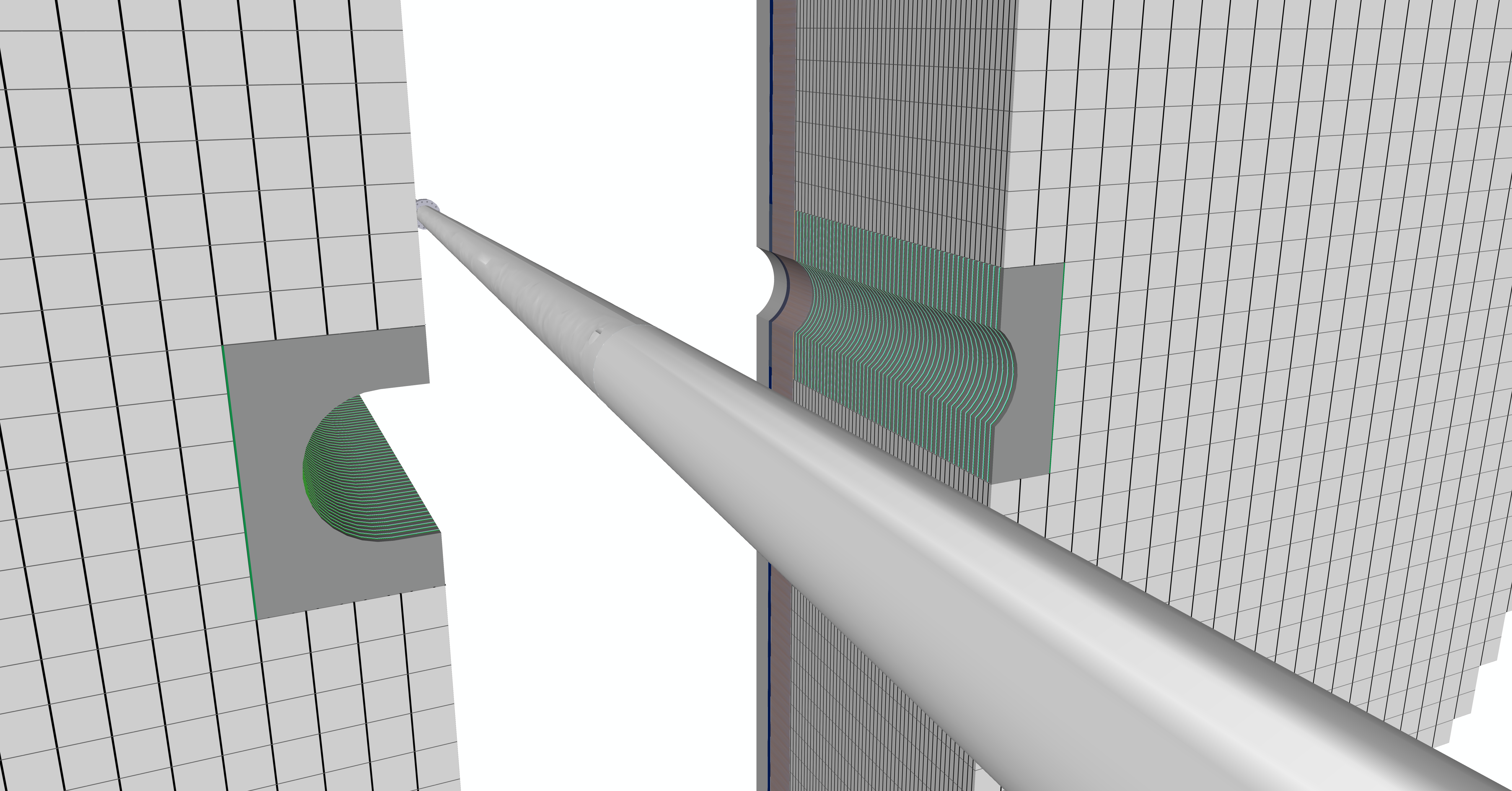}
    \caption{Top: Zoom-in of the downstream face of the HCal, including the insert, in the closed configuration.  Bottom:  Same in the open configuration. The support structure and rail system that will hold the entire endcap is not shown. }
     \label{fig:split}
 \end{figure}

The scintillator assemblies could be removed for maintenance by sliding them out laterally from between the absorbers when the HCal is opened, as illustrated in Fig.~\ref{fig:split}. This will be possible because the interaction point will have a rail system that will be integrated to the support structure~\cite{AbdulKhalek:2021gbh}. 
 
This feature ensures that the SiPMs boards could be annealed at high-temperature after each run to minimize radiation damage, and that SiPMs and scintillator tiles could be replaced.
\FloatBarrier
\subsection{Geometric acceptance}
\label{geometric_acceptance}
The overall effects of the EIC crossing angle and other beam parameters are discussed in Ref.~\cite{crossangle_1}. Events in the lab frame, where the beams have a non-zero crossing angle, can be Lorentz transformed into a frame where the beams are anti-parallel and have energies very close to those in the lab frame. This ``minimally-transformed, collinear frame'' is a natural choice to extract many physics observables and the accompanying particles' angular distributions. 

In the hadron- (electron-) going direction, scattering angles defined in the lab frame relative to the hadron (electron) beam will be equivalent to the scattering angle in the minimally-transformed collinear frame~\cite{crossangle_2}. In particular, for a detector in the hadron-going direction, its angular acceptance should be defined relative to the hadron beam.

\begin{figure}[h]
    \centering
     \includegraphics[width=0.49\textwidth]{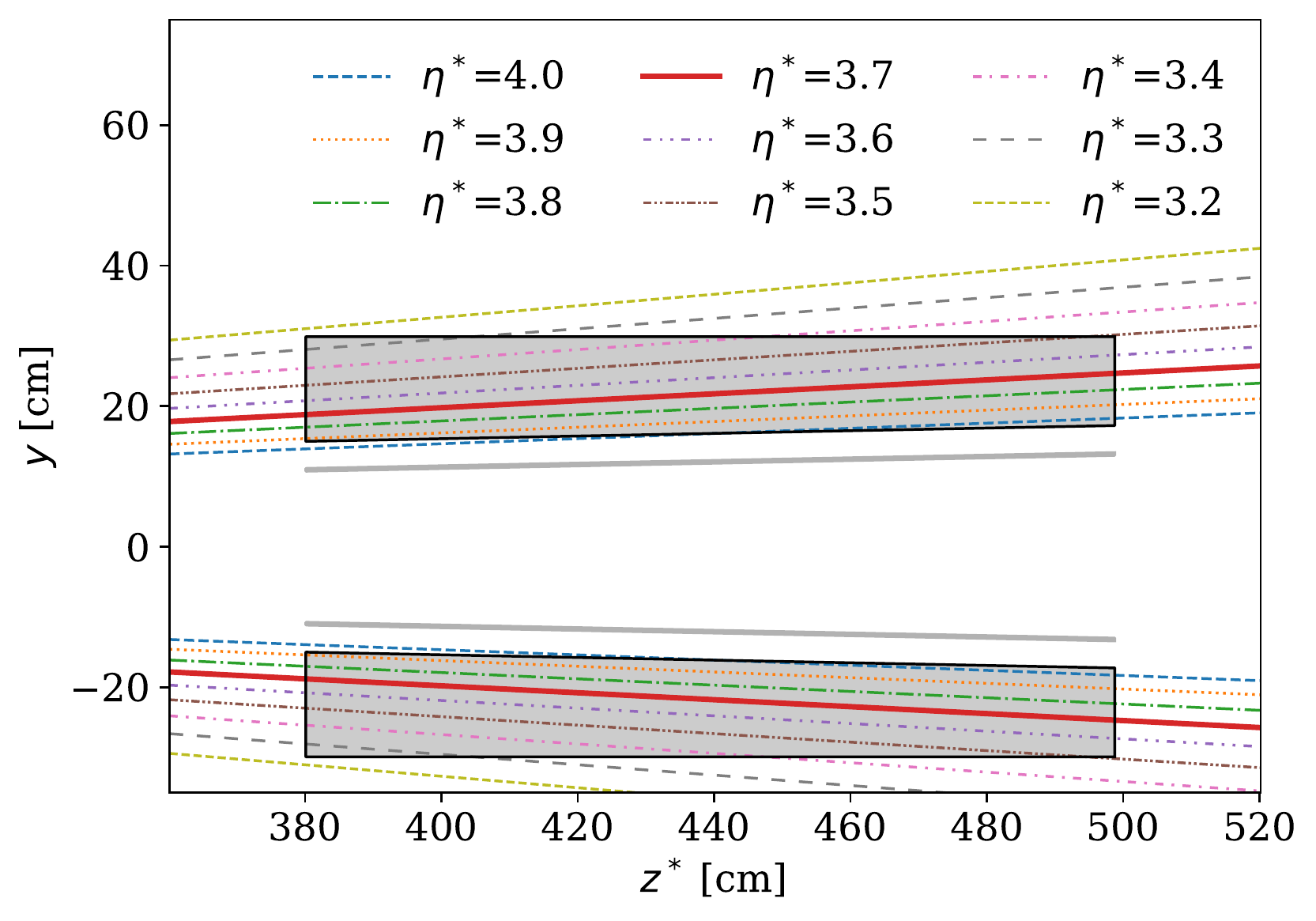}
    \caption{Contours of constant $\eta^{*}$, superimposed on a vertical vs.~longitudinal cross section of the insert.  The thick curve (red online) represents $\eta^*=3.7$.}
   \label{fig:eta_vs_y_z}
\end{figure}

To estimate the geometrical acceptance of the insert, we determined where cones of constant $\eta^*$ (that is, pseudorapidity with respect to the $z^*$ axis -- the proton beam axis) would intersect the detector. 
Figure~\ref{fig:eta_vs_y_z} shows projections of these $\eta^*$ cones superimposed on a $yz^{*}$ (vertical vs.~longitudinal) cross-section of the insert. Within this slice, the $\eta^*=4.0$ cone skims the inner edge of the detector, while those at lower $\eta^*$ hit the face of the detector.  

\begin{figure*}[h]
    \centering
    \begin{overpic}[width=\textwidth]{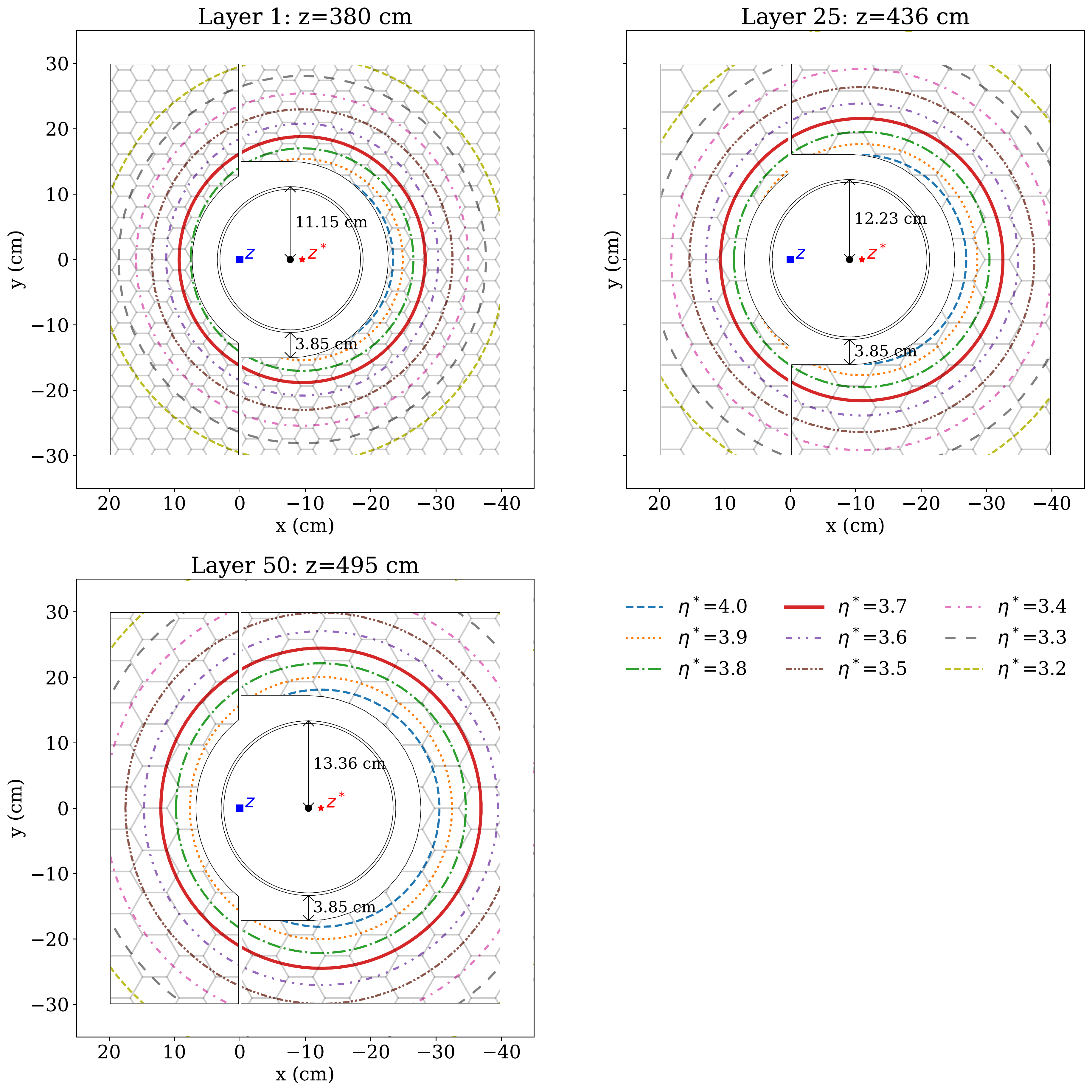}
    \put (55, 5) { \includegraphics[width=0.45\textwidth]{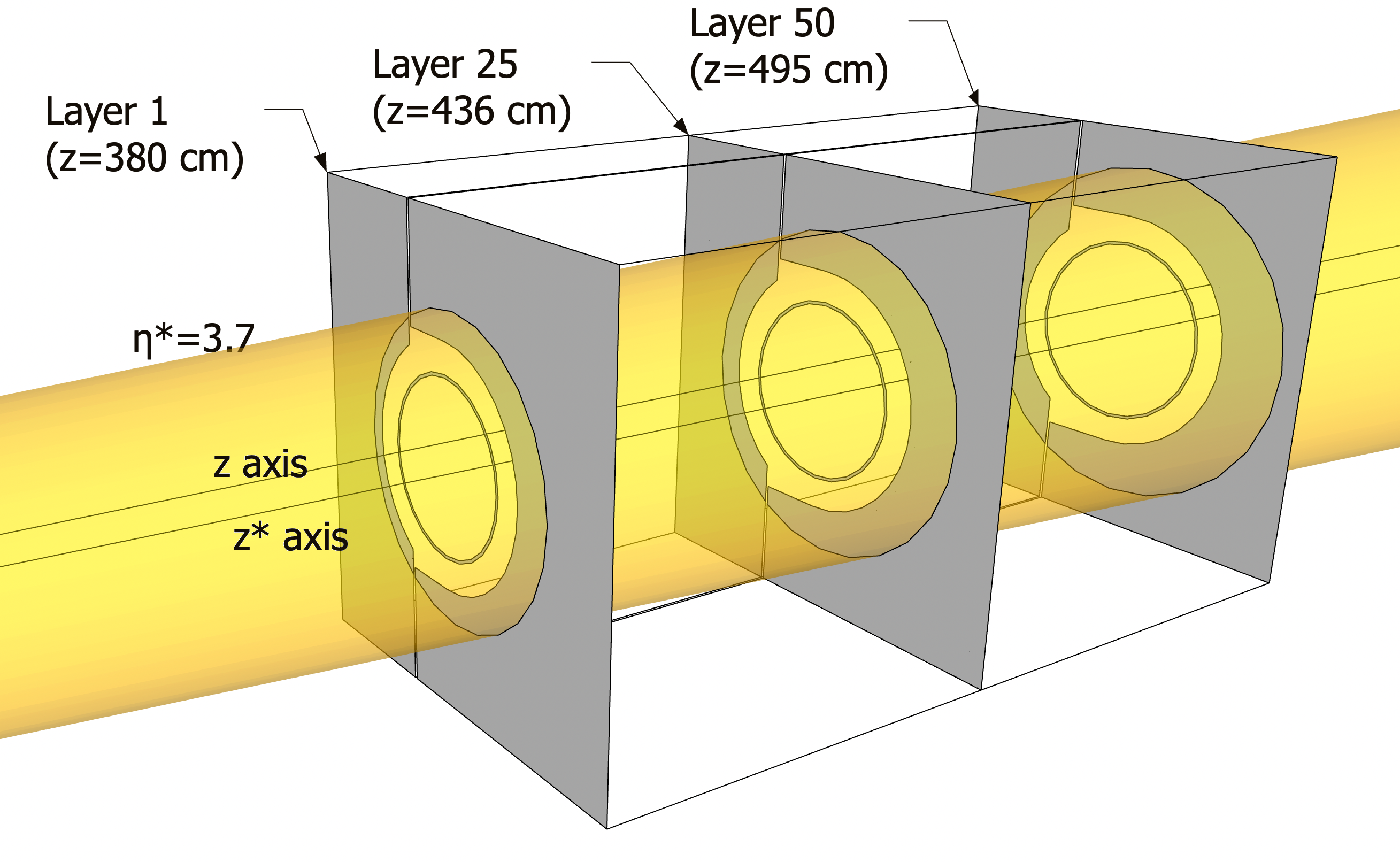}}
    \end{overpic}
    \caption{Contours of constant pseudorapidity relative to the $z^*$ axis, $\eta^{*}$, in the first layer (top left), one of the middle layer (top right), and the last layer (bottom left) of the insert, using the baseline value of the clearance (3.85~cm). The scintillator cell boundaries are superimposed. Inset: Example cone of particles at $\eta^*$=3.7 (corresponding to the thick red ring on the other panels) projected into the calorimeter insert, showing cross-sections of the cone at the first, middle, and last scintillator layers.}
    \label{fig:clearance}
\end{figure*}

To determine the geometrical acceptance over the full azimuthal range we projected these cones of constant $\eta^*$ onto the faces of the scintillators planes of the insert, resulting in elliptical rings of constant $\eta^*$. An example is shown in the inset of Fig.~\ref{fig:clearance} for $\eta^*$=3.7, which covers the full azimuthal acceptance on all layers of the detector.

We show in Fig.~\ref{fig:clearance} the cross sections of these cones for several values of $\eta^*$ at the first layer (top left panel), one of the middle layers (top right panel), and the last layer (bottom left panel). 
While there is some acceptance up to  $\eta^*=4.0$, this is limited to one side of the detector, and it barely skims the first layer, causing a degradation of the signal, as we quantify in Sec.~\ref{sec:simulation}.
 
Since the final detector design may differ from the baseline design presented in this work, we discuss possible alternatives: with a clearance parameter of 3~cm instead of 4~cm, the entire $\eta^*=3.8$ ring fits entirely within the first scintillator layer. If the clearance parameter were increased to 5~cm, the full $\eta^*=3.7$ ring would still be within the acceptance of the detector, but it would come very close to the edge, resulting in some degradation.

The relevant scale to compare various clearance options has to be the effective radiation length of the insert, which in our design is minimized using mostly tungsten as an absorber. This is because the relevant metric is whether or not the core of the hadronic showers can be well contained, as we describe more in Sec.~\ref{sec:simulation}. 

Given that even seemingly small differences of clearance definitions can have a large impact in extending the $\eta^*$ coverage of the detector, careful engineering design and studies are required to minimize the clearance as much as possible.

\subsection{Total calorimeter thickness}

\begin{figure*}[h!]
    \centering
    \includegraphics[width=0.9\columnwidth]{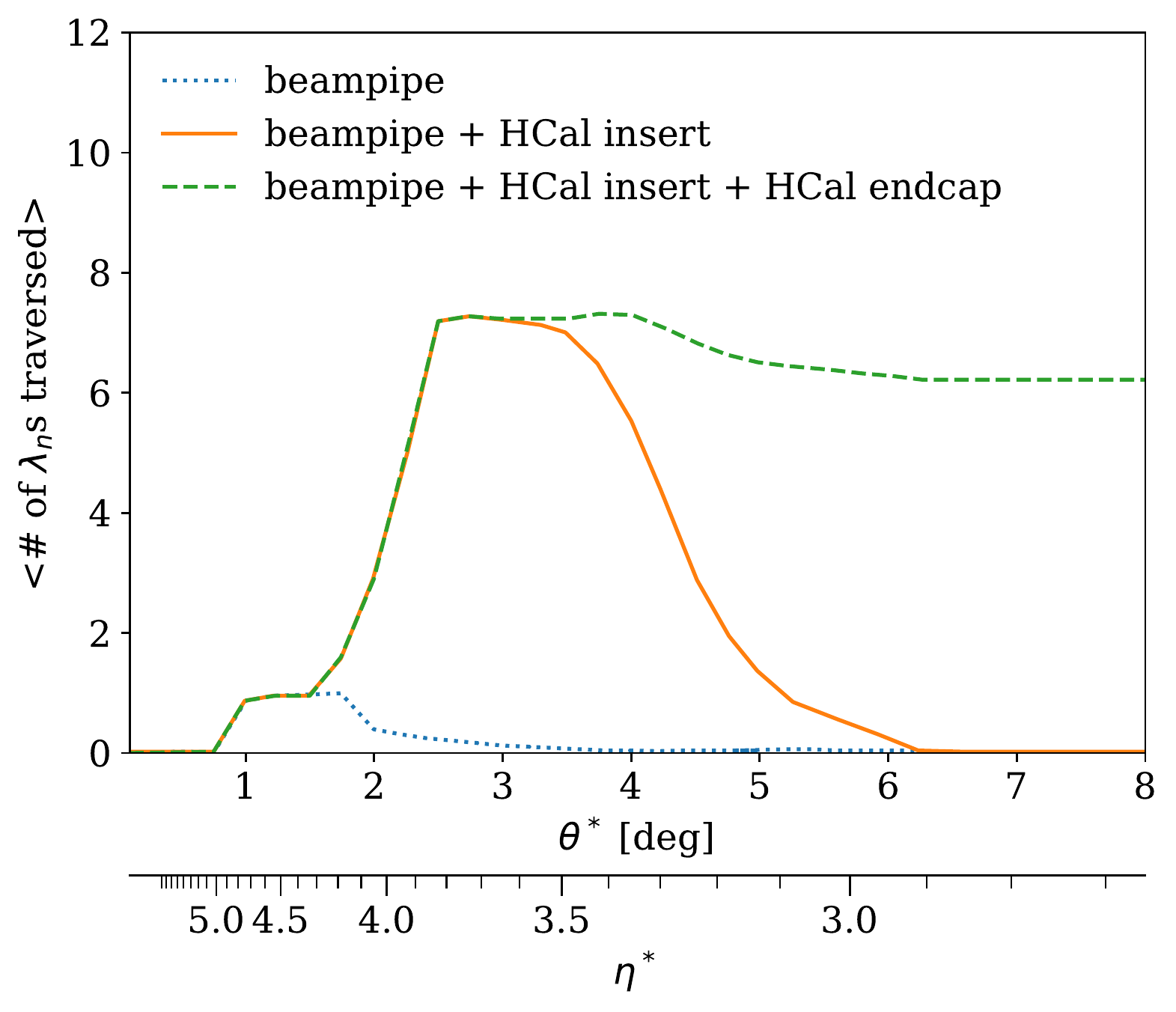}
    \includegraphics[width=\columnwidth]{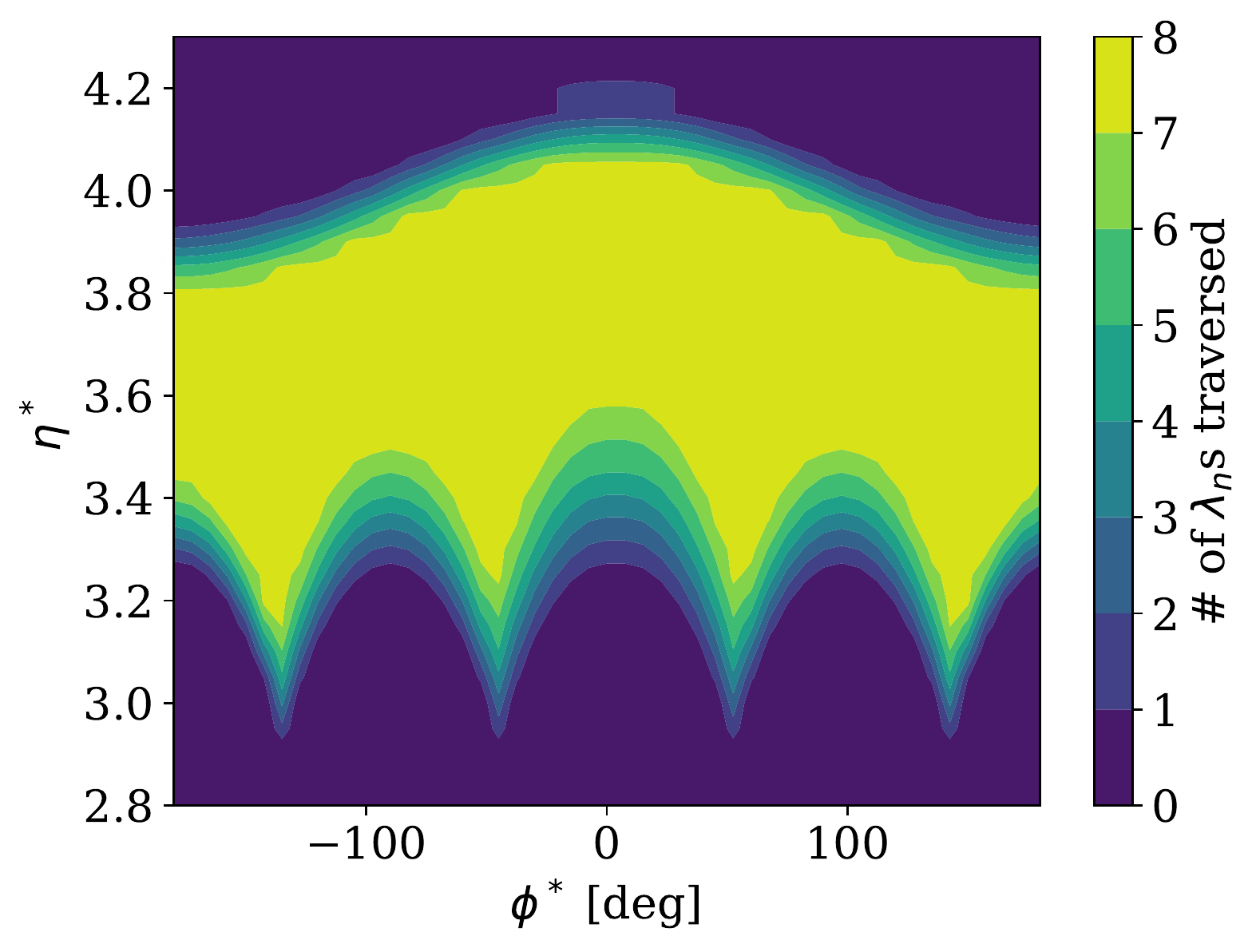}
    \caption{Left: Number of nuclear interaction lengths of material traversed by a particle moving in a straight line from the origin as a function of $\theta^*$.  These values are integrated over $\phi^*$.  The different colors represent the material in the beampipe only (dotted curve, blue online), the beampipe and the HCal insert (solid curve, orange online), the beampipe + the proposed HCal insert + the HCal endcap (dashed curve, green online). Right: Number of nuclear interaction lengths of material in the HCal insert traversed by a particle moving in a straight line from the origin, as a function of both $\phi^*$ and $\eta^*$.}
    \label{fig:nuclear_interaction_lengths}
\end{figure*}

We determined the amount of material (in nuclear interaction lengths, $\lambda_n$) that a particle travelling in a straight line would pass through in the HG-CALI and the beampipe as modeled in Ref.~\cite{beampipe}, as a function of its polar angle $\theta^*$ with respect to the $z^*$ axis. We averaged this with respect to the azimuthal angle $\phi^*$ and show the results in the left panel of Fig.~\ref{fig:nuclear_interaction_lengths}.

The HG-CALI covers polar angles from about 2$^\circ$ to 6$^\circ$, with up to about 7 nuclear interaction lengths. Between 3$^\circ$ and 6$^\circ$, there is overlap between the HG-CALI and the HCal endcap. The HCal endcap has a somewhat lower number of nuclear interaction lengths (about 6) than the insert.  There is a small amount of material traversed by particles passing through the sides of the beampipe (about 1$\lambda_n$), which may allow the detection of showering at very small $\theta^*$.

We show in the right panel of Fig.~\ref{fig:nuclear_interaction_lengths} the dependence of the amount of material traversed in the HCal insert on $\phi^*$ and $\eta^*$.  The amount of material is nearly uniform in $\phi^*$ in the $3.5\leq \eta^*\leq 3.8$ range. Since the insert's outer shape is (approximately) a square, the value of $\eta^*$ at the edge of the insert depends non-trivially on $\phi^*$. At about $3.0<\eta^*<3.3$ (depending on $\phi^*$) particles may skim the edge of the insert, creating a signal in both the insert and the HCal. Particles with $3.3<\eta^*<3.9$ encounter at least two nuclear interaction lengths of material in the insert at all azimuthal angles.

The HG-CALI can complement the HCal endcap at angles where the effective thickness is not maximal by providing a measurement of the shower-start position and incident-particle angle, both of which will be better measured in the insert.

%% file: Simulations.tex
\section{Simulation and performance}
\label{sec:simulation}

\subsection{Geometry and simulation setup with \textsc{DD4HEP}}
We used the \textsc{DD4HEP} framework~\cite{Frank:2014zya} to define the geometry and ran the \textsc{Geant4} framework~\cite{GEANT4:2002zbu} to evaluate the performance of the HG-CALI. The version of \textsc{Geant4} used is 10.7.p03. 

Table~\ref{tab:layer} shows the materials, thicknesses, and numbers of interaction lengths for each component of a single layer of the insert (see Fig.~\ref{fig:explode_view}). The previously described layer design is followed: 30 layers W/Sc, 20 layers steel/Sc, and 1 final layer of steel.

For simplicity, the granularity of the insert was assumed to be 3$\times$3~cm$^{2}$ for all the layers. Additionally, the model did not include the dimples and grooves in the scintillator layers, nor the SiPMs in the PCB. The model included a realistic beampipe geometry and material description~\cite{beampipe}.

We also included in the simulation an HCal model similar to the STAR/ATHENA design~\cite{Tsai:2015bna,ATHENA}, assuming a granularity of 10$\times$10~cm$^{2}$, 20~mm thick absorbers, and 3~mm thick scintillators. 

\begin{table}[h]
    \centering
        \caption{Summary of a single layer of sampling calorimeter structure. Separate values are given for layers with tungsten absorbers and those with steel absorbers.}
    \begin{tabular}{c|c|c|c}
        Material            & density  & thickness &
        $n_\lambda$\\ 
        & (g/cm$^3$) & (cm)\\
        \hline
        Absorber (tungsten) & 19.3 &  1.61 & 0.15566 \\
        Absorber (steel) & 7.85 &  1.61 & 0.09364 \\
        Air & 0.0012 & 0.02 & $3\times 10^{-7}$ \\ 
        Aluminum & 2.699 & 0.20 &     0.00502 \\ 
        ESR & 1.06 & 0.015 &          0.00022\\
        Scintillator &  1.06 & 0.30 & 0.00436\\ 
        ESR & 1.06 & 0.015 & 0.00022\\
        PCB & 1.86 & 0.16 & 0.00352\\ 
        Air & 0.0012  & 0.02 & $3\times 10^{-7}$ \\ 
        \hline
        Total (tungsten) &  --  & 2.34 & 0.16899 \\
        Total (steel) & --   & 2.34 & 0.10697 \\
    \end{tabular}
    \label{tab:layer}
\end{table}
\FloatBarrier
The solenoidal magnetic field is expected to  be low in the forward region relevant for the insert, but is not precisely known at this time so it was not included in the simulation.

For the simulation of the particle shower in the material, we used the \textsc{FTFP\_BERT\_HP} physics list of \textsc{Geant4}, which well describes the CALICE test-beam data~\cite{CALICE:2015fpv}, including the time evolution of hadronic showers in both tungsten and steel absorbers~\cite{CALICE:2014tgv}. To model saturation effects in the scintillator response we applied Birks' law \cite{Birks:1951boa} with a constant 0.126~mm/MeV.

The particle origin position was assumed to be at $z=0$ and the first layer of the HCal insert was placed at $z=380$~mm. 

No noise was included in the simulation, as a realistic implementation is not available at this time, but its effect was accounted for with an energy threshold as described in the next section. For the digitization step, a 12-bit dynamic range for the ADC and maximum value set to 200~MeV. 

Each cell was assigned a timestamp dictated by the particle with the earliest time of arrival to the cell. The effect of the time resolution was included as a cell-by-cell smearing with a Gaussian with $\mu=0$, $\sigma=$1~ns. Unlike in Ref.~\cite{CALICE:2015fpv}, no SiPM-saturation effect was simulated since this effect is expected to be negligible with the model chosen as baseline\footnote{For reference, the SiPM used in the CALICE W-AHCAL prototype contained about 1k channels whereas the one used in our design has 14.4k channels.}. 

The hits were also required to have $E_{\mathrm{hit}} > 0.1E_{\mathrm{MIP}}$, where $E_{\mathrm{MIP}}$ is the most-probable energy deposited by a minimum-ionizing particle (MIP), which is about 0.8~MeV as obtained from simulations of high-energy muons. This energy cut was assumed to reduce the noise to a negligible level, as it will correspond to a cut of about 6~photoelectrons. Our selection was more relaxed than the CALICE studies~\cite{CALICE:2010fpb} that used an older generation of SiPMs with significantly larger noise, after-pulsing, and cross-talk.

We used a time cut of $t<50$~ns, where $t=0$ is defined when the generated particle hit the front of the insert. As we will show in Sec.~\ref{sec:linearity}, this selection yields a compensated response with a basic reconstruction algorithm. Much larger integration times might be possible given the EIC bunch structure and absence of pileup. Such time information can be exploited with advanced algorithms, such as those proposed in Refs.~\cite{Graf:2022lwa,Akchurin:2021afn,Qasim:2021hex}.
\subsection{Shower-shape examples}
\label{sec:showershape}
\begin{figure*}[h]
    \centering
  \includegraphics[width=.32\textwidth]{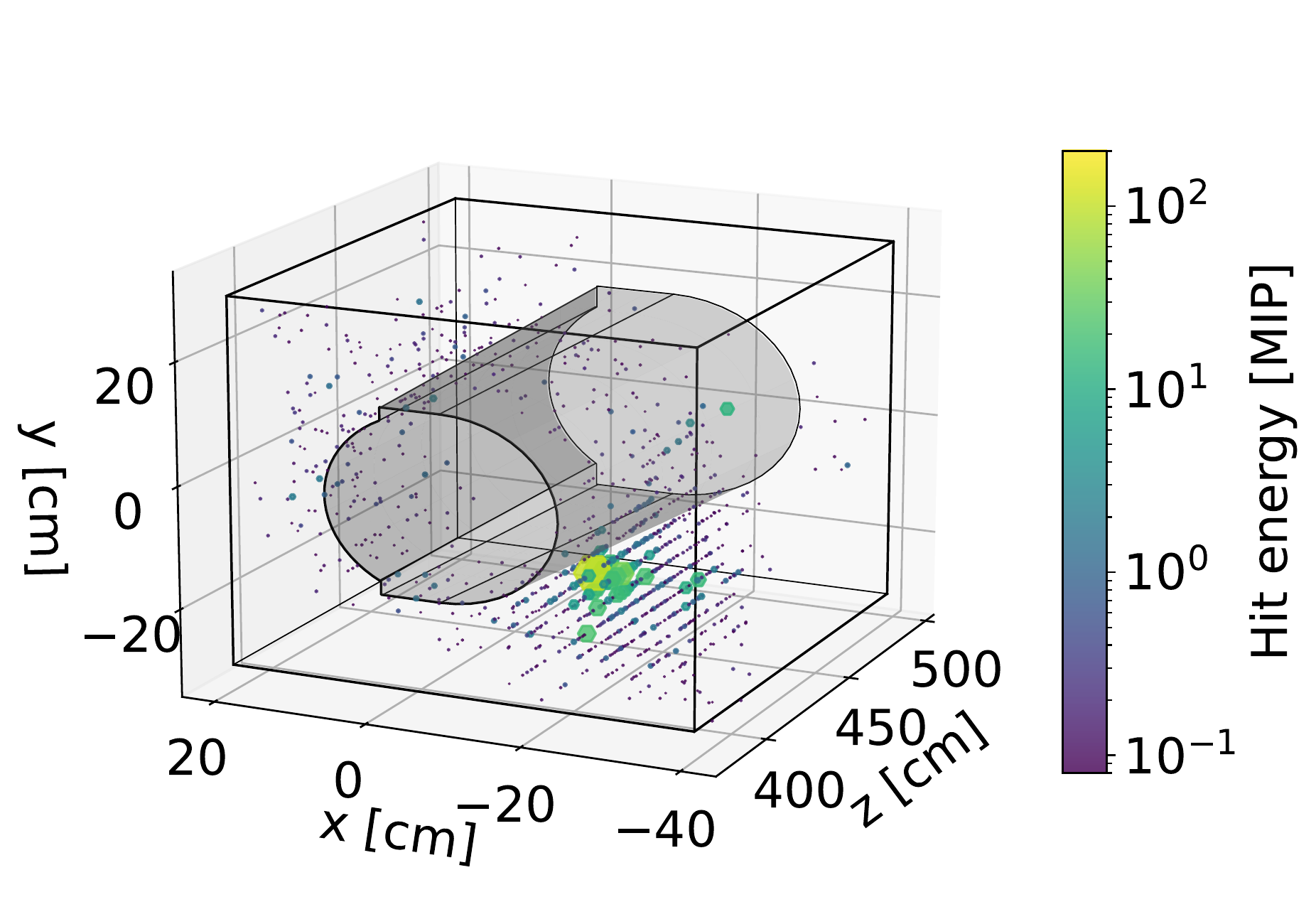}\includegraphics[width=.32\textwidth]{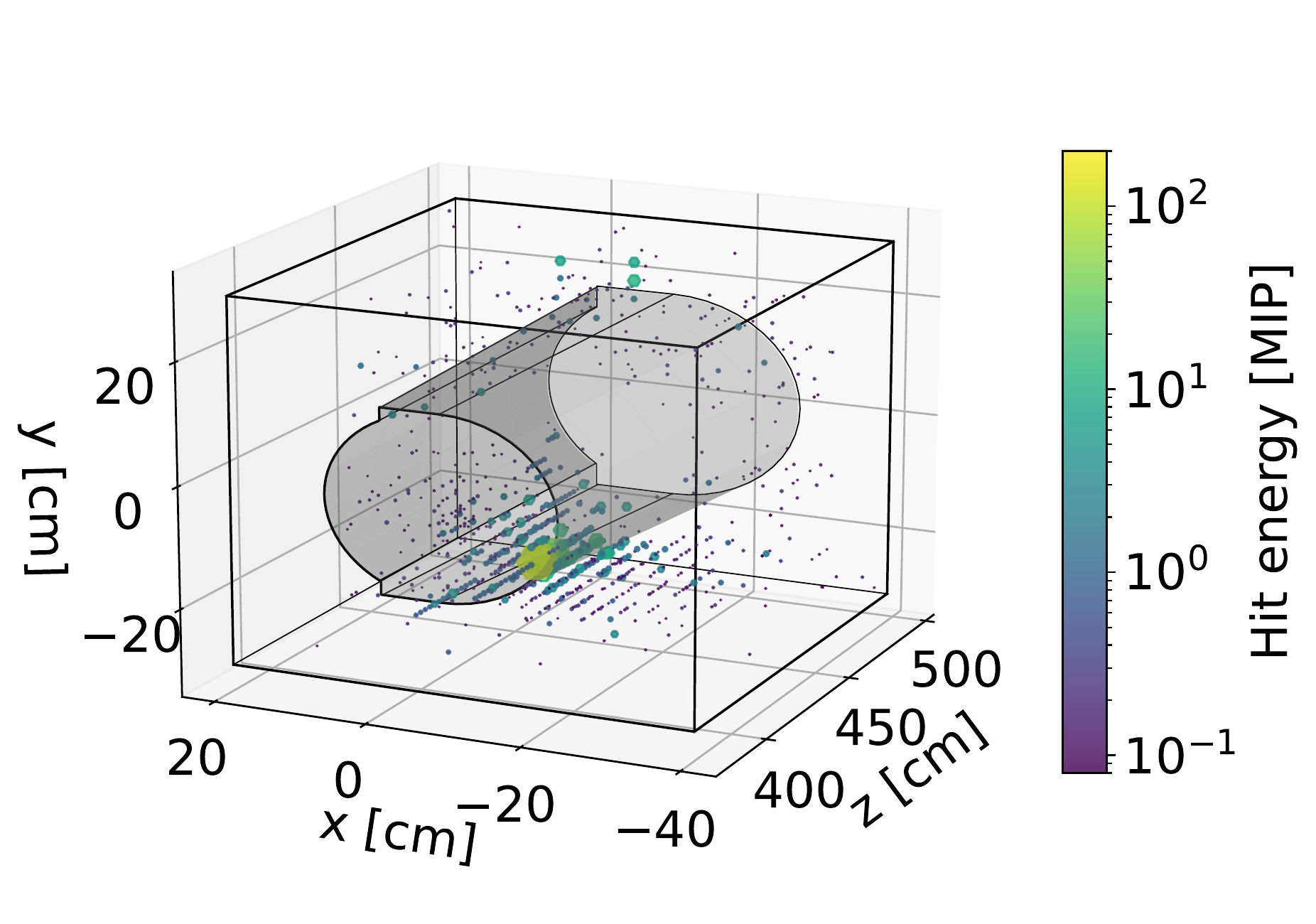}\includegraphics[width=.32\textwidth]{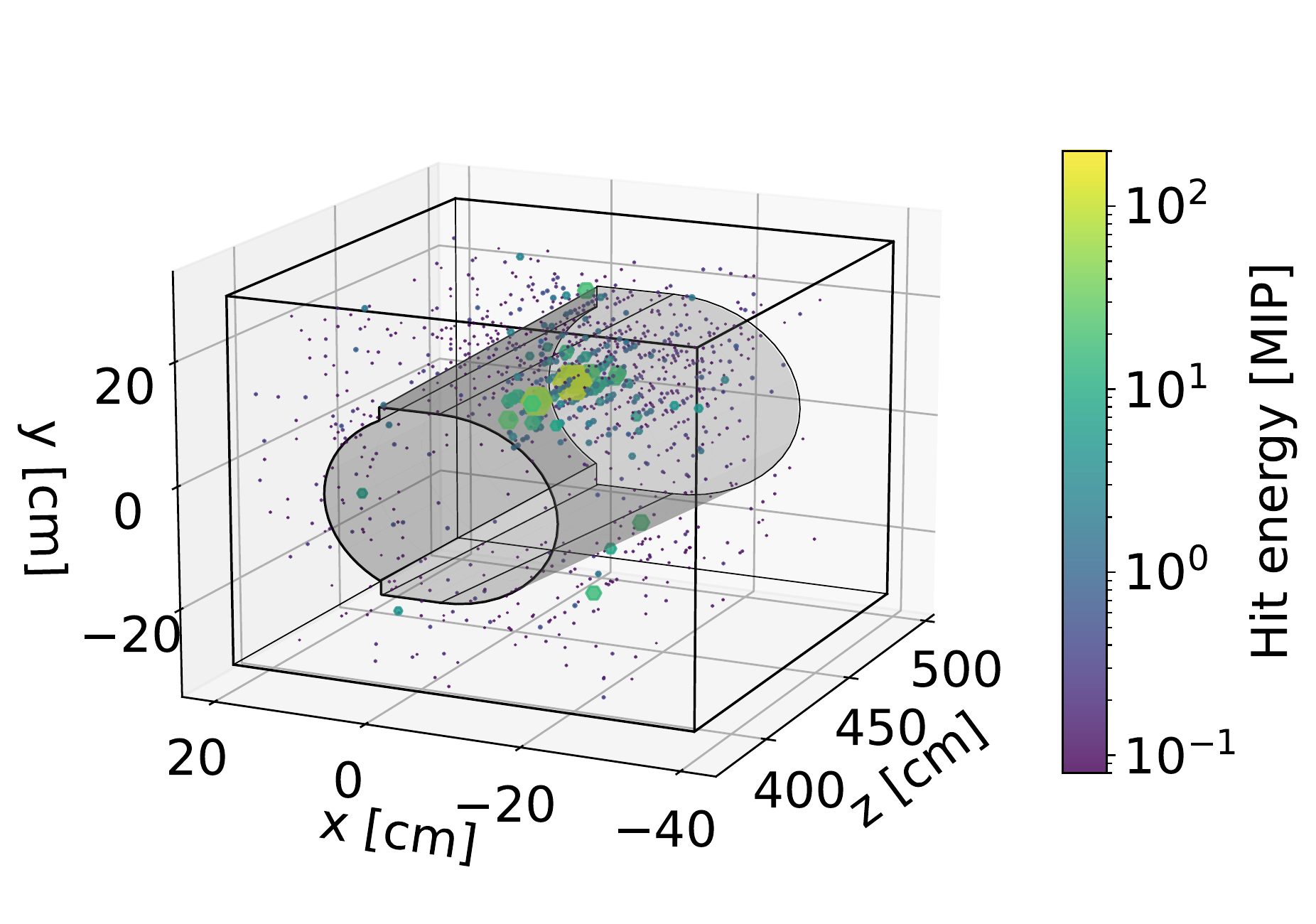}\\
  \includegraphics[width=.32\textwidth]{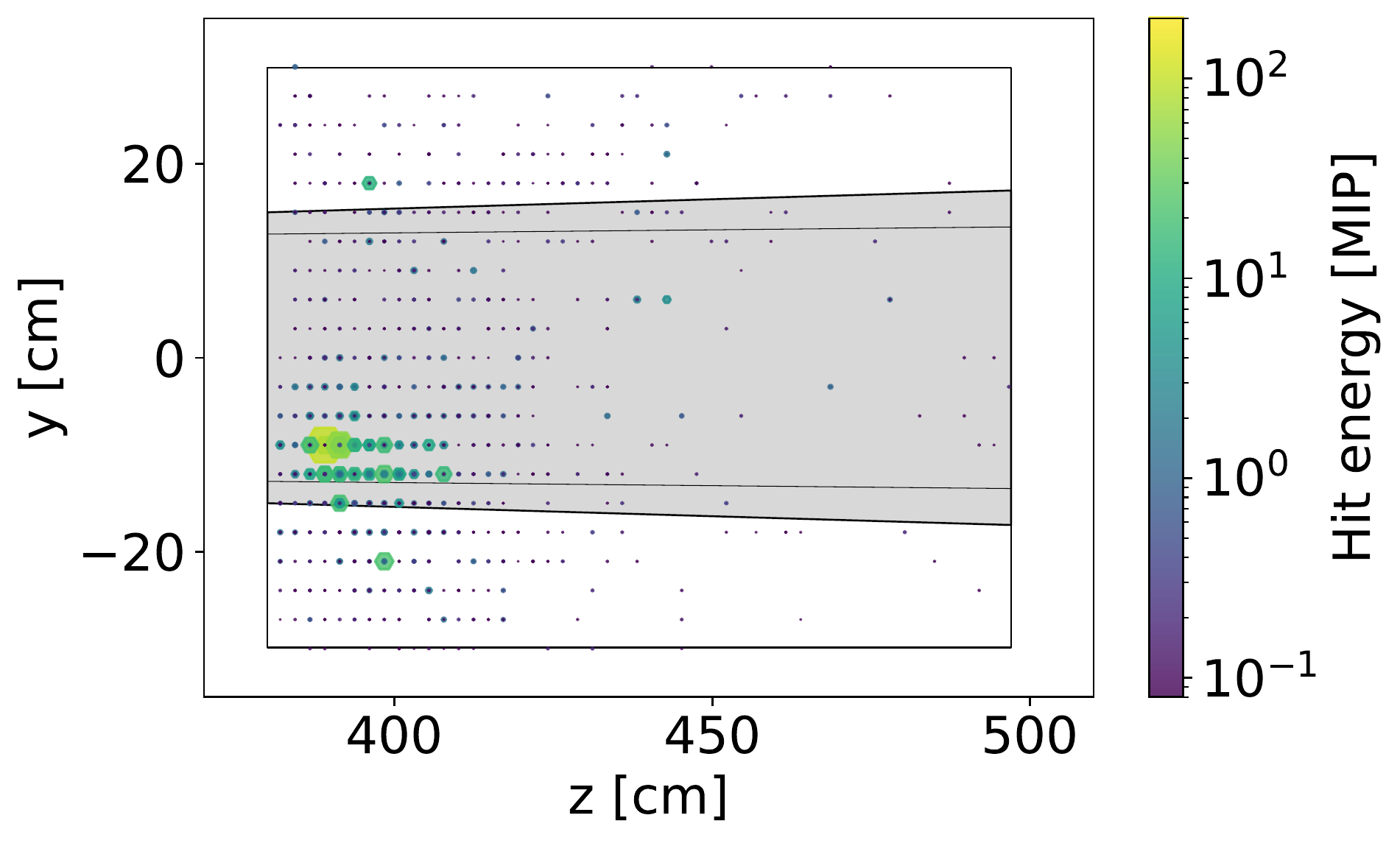}\includegraphics[width=.32\textwidth]{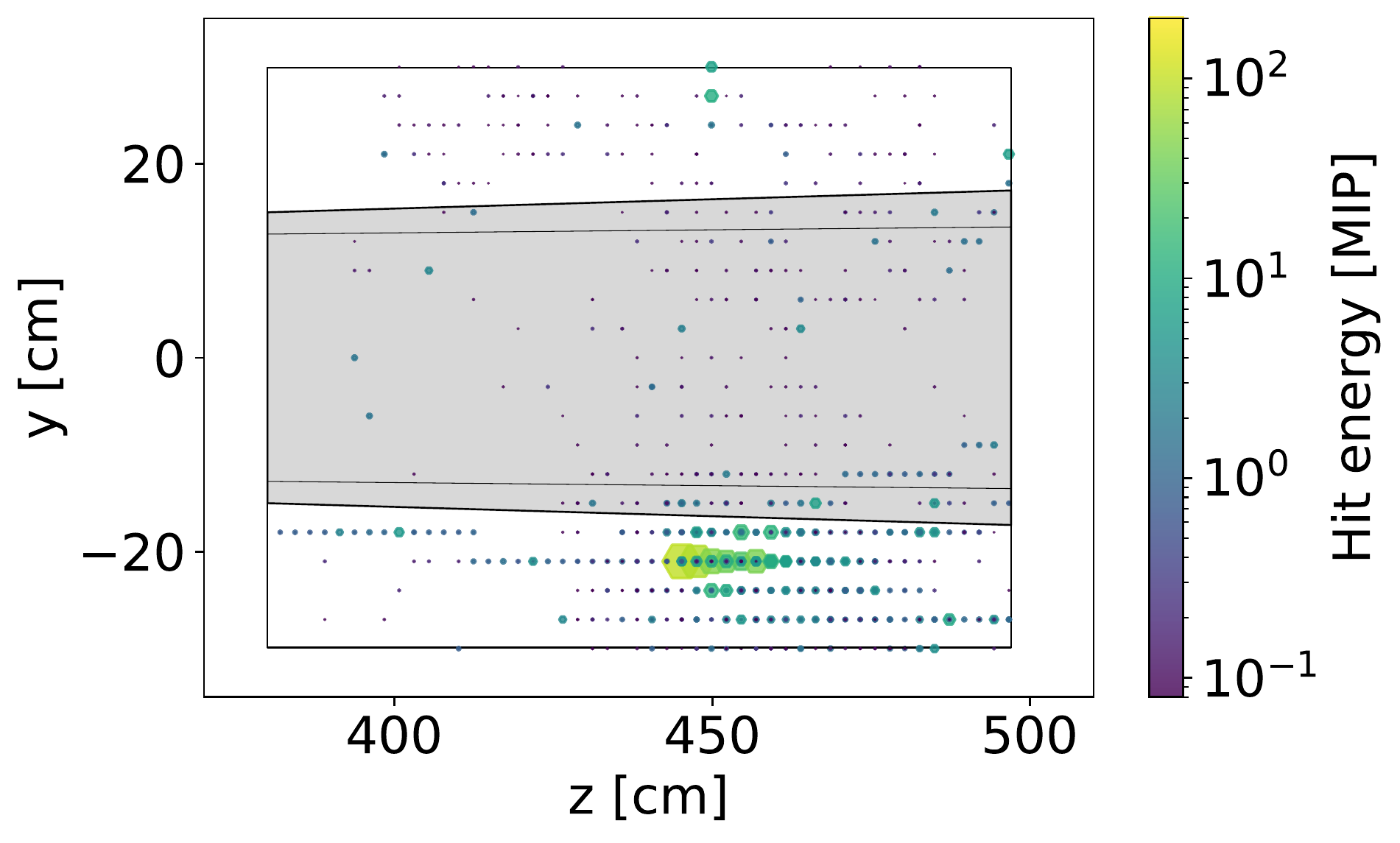}\includegraphics[width=.32\textwidth]{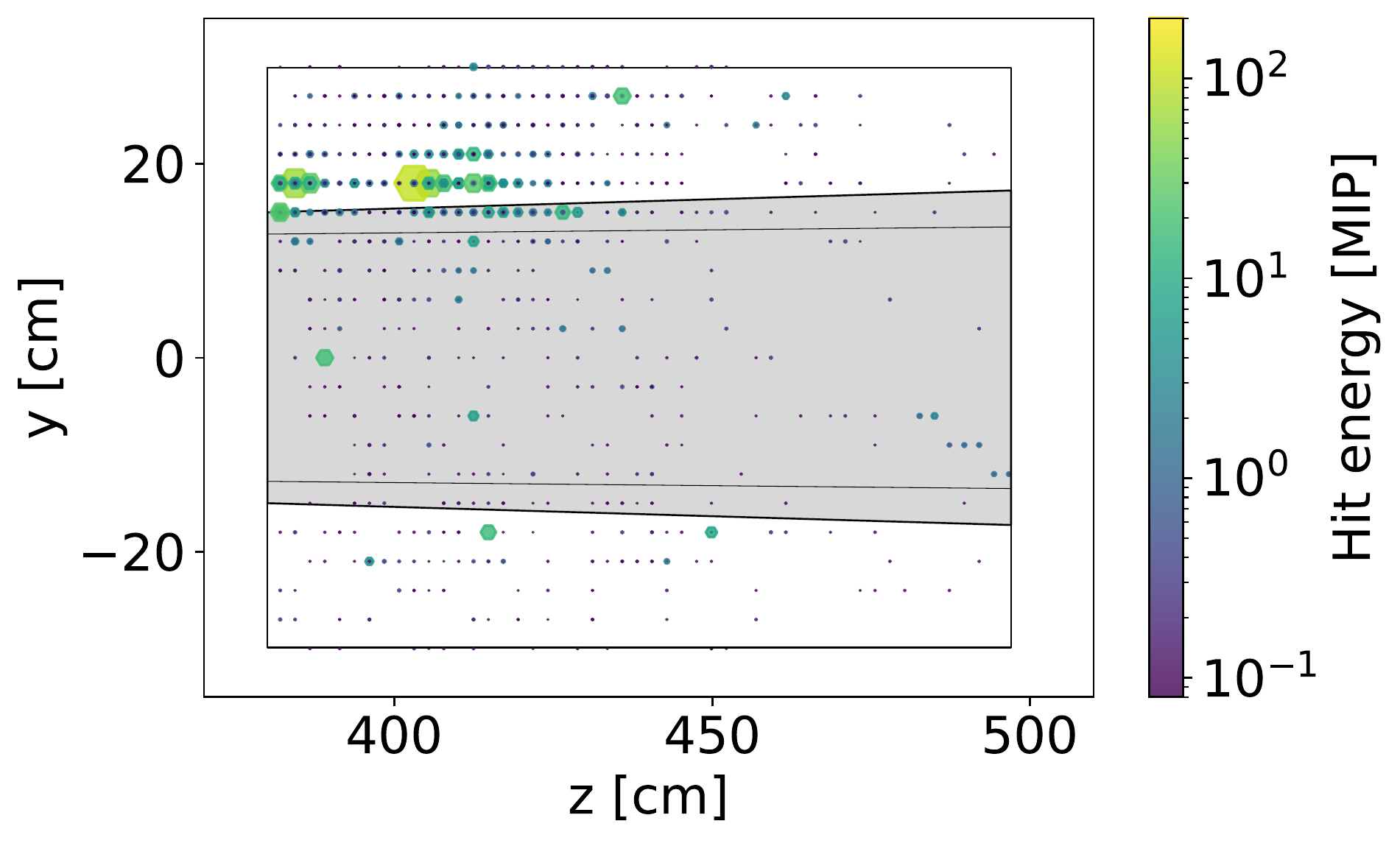}
    \caption{Examples of reconstructed shower shapes in HCal insert for 50 GeV $\pi^{-}$ at $\eta^{*}=3.7$. Threshold and timing cuts are applied, as described in the text. The color code and size of the marker shown represent the hit energy. A wireframe shows the boundaries of the HCal insert, and the hole is shown in grey. In the top row the showers are shown in 3D, and in the bottom row these are projected in 2D for the same set of events.}
    \label{fig:showershape}
\end{figure*}

Figure~\ref{fig:showershape} shows some example shower shapes for a 50~GeV $\pi^{-}$ with $\eta^{*}=3.7$ and a random $\phi^{*}$. As expected, the $\pi^{-}$ showers can be roughly described as having a narrow core from electromagnetic sub-showers and a halo from hadronic sub-showers. The core typically contains cells with large energy deposits, whereas the halo is composed of mainly low-energy cells. The latter are dominated by the copious production of neutrons produced in tungsten. Similar observations have been made with CALICE prototypes~\cite{Sefkow:2015hna}.

The hadronic-shower shapes, although irregular, provide a hint on the performance of the insert. The production of neutrons, which drives the hadronic halo, occurs over wide angles. Some of the neutrons will be emitted in the direction of the beampipe hole, but often these appear in the other side of the calorimeter instead of being lost altogether. On the other hand, the core of the shower is electromagnetic in nature so its size is driven by the radiation length, which is short in tungsten.

\subsection{Hit distributions}

\begin{figure}[h!]
    \centering
    \includegraphics[width=\columnwidth]{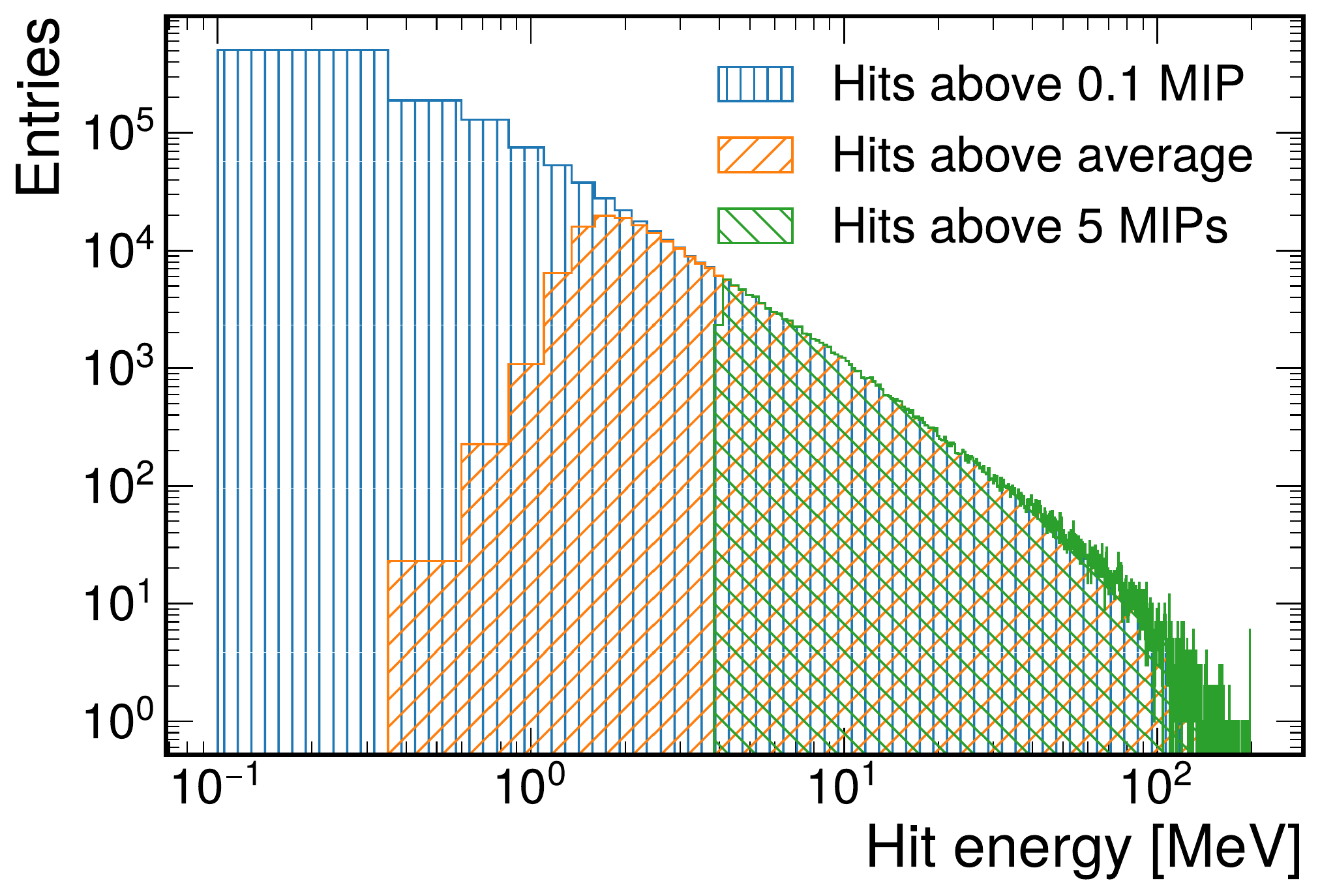}\\
        \includegraphics[width=\columnwidth]{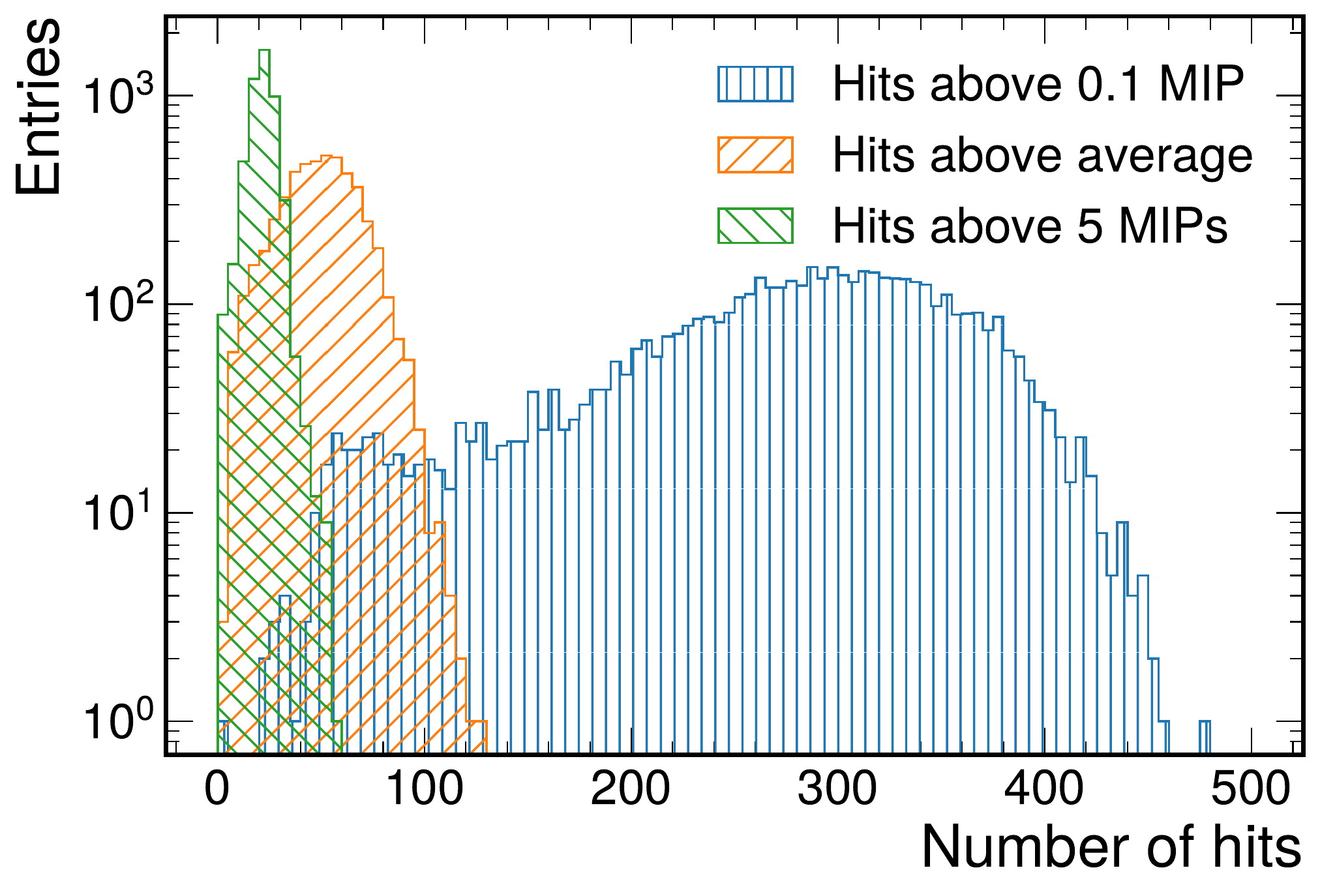}
    \caption{Reconstructed hit distributions for 50 GeV pions at $\eta^{*}$=3.7 with different hit energy thresholds. Top: Hit energy distribution. Bottom: Hit multiplicity. }
    \label{fig:hitdistributions}
\end{figure}

Figure~\ref{fig:hitdistributions} shows the hit-energy and multiplicity distributions for a 50~GeV $\pi^{-}$ with $\eta^{*}=3.7$. We show the distributions separately for all hits above the 0.1~MIP threshold, all hits that have more energy than the average of all the hits in their respective showers, and those with more than 5~MIPs of energy.

The average hit energy was found to be about 1.6~MeV, with 17\% of the hits containing energy above 1.6~MeV and 7.6\% containing energy above 5~MIPs (4~MeV). 
The maximum hit energy was about {200~MeV}. The hits-per-shower distribution shows a broad peak around 275 hits, whereas the hits above average show a narrower distribution centered around 50 hits, and those above 5~MIPs show an average of 20~hits. 

\subsection{Linearity for hadrons and electrons and $e/h$ ratio}
\label{sec:linearity}

\begin{figure*}[h!]
    \centering
     \includegraphics[width=\textwidth]{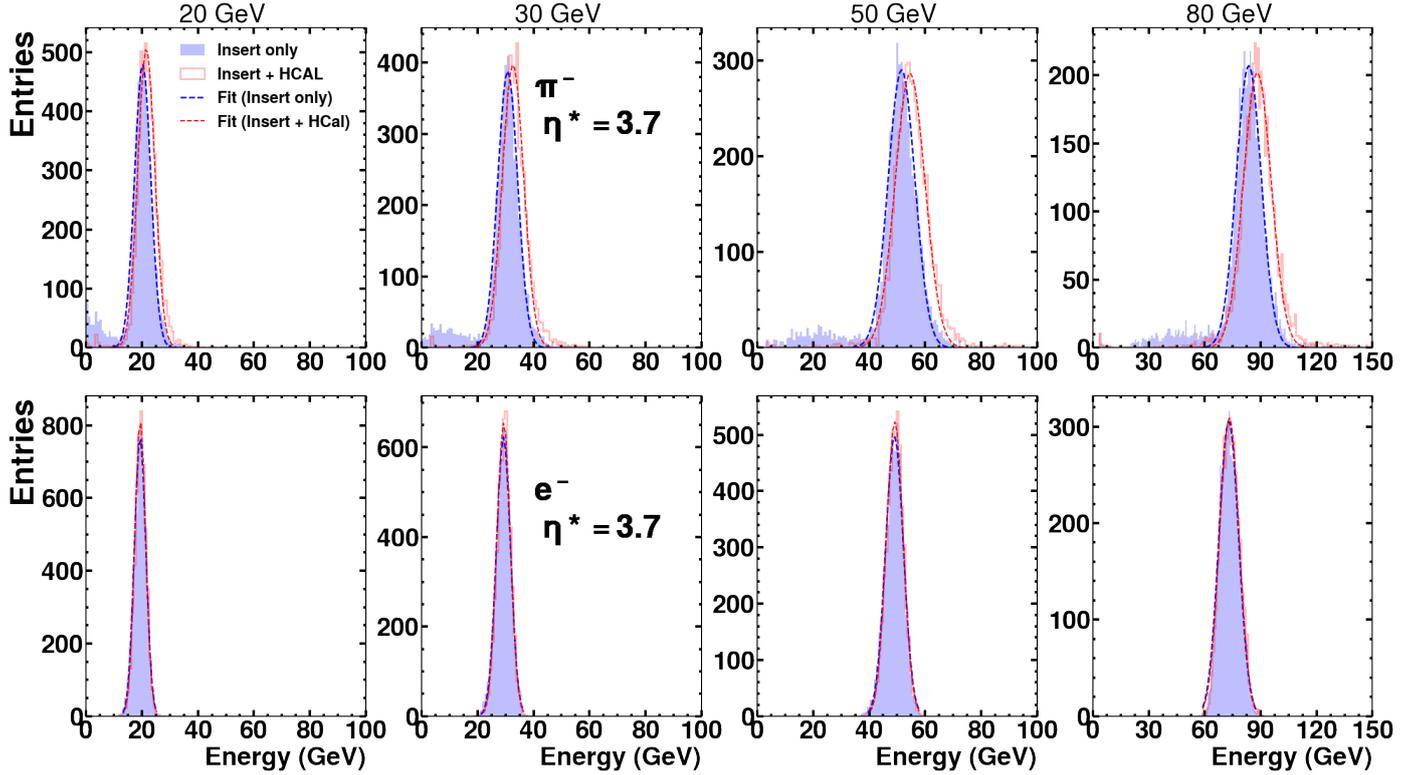}
        \caption{Sum of the hit energies for $\pi^-$ (top row) and $e^-$ (bottom row) with different energies and $\eta^{*}=3.7$. The filled histogram (blue online) is the sum of the energies of the hits from the HCal insert while the outlined one (red online) is the sum of energies of hits from the HCal and the HCal insert, weighted by their respective sampling fractions. The dashed curves represent Gaussian fits to the energy response of the respective distribution. The fits are restricted to the range $\pm2.5$ $\sigma$ from the peak with an iterative procedure.}
    \label{fig:response}
\end{figure*}

 Figure~\ref{fig:response} shows the energy distributions of showers at $\eta^{*}$= 3.7 both for $\pi^-$ and $e^-$ with various energies. The total reconstructed energy was computed as:
 \begin{equation}
 \label{eqn_wt_sum}
     E_{total}=\frac{E_{Insert}}{SF_{Insert}} + \frac{E_{HCal}}{SF_{HCal}}
 \end{equation}
where $E_{Insert}$ and $E_{HCal}$ are the sum of hit energies in the insert and HCal respectively, and $SF_{Insert}$ and $SF_{HCal}$ are the sampling fractions for the insert and HCal, respectively. The sampling fraction is computed by taking the ratio of the output energy to the input energy in the calorimeter. The sampling fraction is 0.9$\%$ for the insert and 2.2$\%$ for the HCal.

 The response for pions in the insert is mostly Gaussian but it shows a low-energy tail, which is much less pronounced with the combined measurement using the HCal insert and the HCal (this is quantified in Sec.~\ref{sec:leakage}). This suggests that the low-energy tail is due to leakage in the transverse direction from the HCal insert towards the HCal. 
 
 The response for electrons is mostly Gaussian. This also suggests that that the electromagnetic core of the hadronic showers is well contained due to short radiation length of tungsten.

\begin{figure}[h!]
    \centering
        \includegraphics[width=\columnwidth]{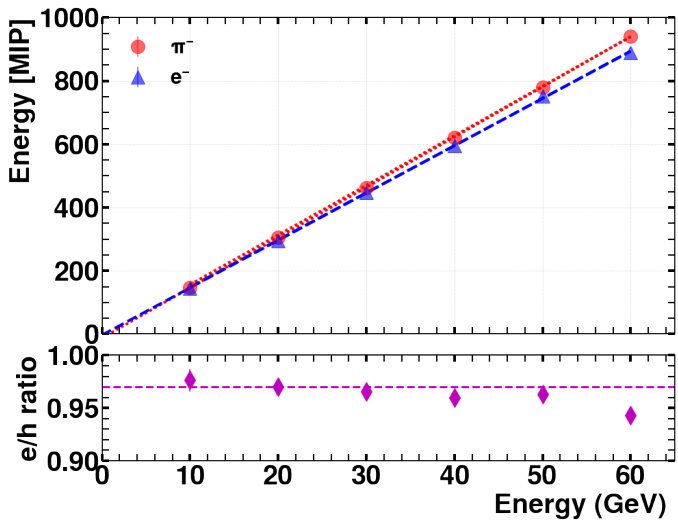}
    \caption{Top: Mean energy of a shower for electrons (triangles, blue online) and pions (circles, red online) at different energies. The dashed and dotted lines (blue and red online) are linear fits for electrons and pions, respectively. Bottom: Ratio of the mean shower energies deposited for electrons to that of pions. }
    \label{ehratio}
\end{figure}

Figure~\ref{ehratio} shows the mean energy deposited in the HCal insert for electrons and pions, along with the $e/h$ ratio, which is found to be about 0.97 on average.

\subsection{Non-Gaussian tails due to leakage}
\label{sec:leakage}

\begin{figure}[h]
    \centering
    \includegraphics[width=0.5\textwidth]{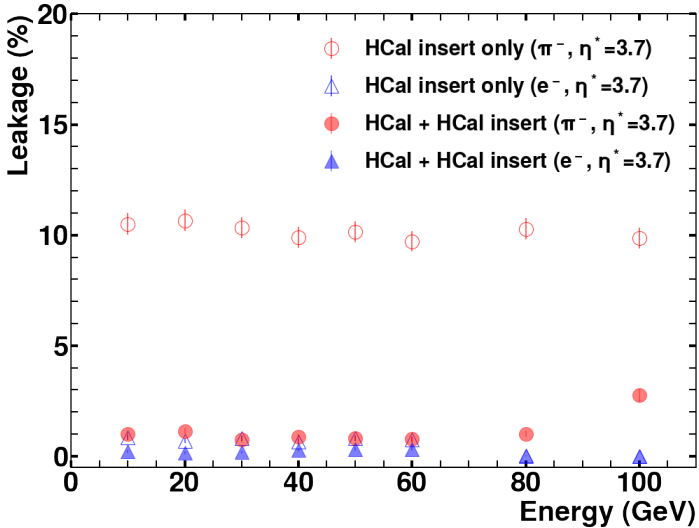}
      \caption{Leakage of shower for $\pi^-$ (red circles) and electron (blue triangles) as a function of energy for $\eta^*$=3.7 (colored online). The open markers represent the analysis done considering the HCal insert only, while the solid markers represent the case where both the HCal and the HCal insert are taken into account.}
    \label{Leakage}
\end{figure}

Leakage is quantified as the fraction of events in which the shower energy is below $\mu-3\sigma$, where $\mu$ and $\sigma$ are obtained from the Gaussian fit to the distribution. Figure~\ref{Leakage} shows the leakage for $\pi^-$ and $e^-$ at different energies and $\eta^*$=3.7. The leakage in the insert alone for $\pi^-$ is about 10\% for the entire energy range but relatively small (about 1\% or lower) for electrons. After considering both the HCal and the HCal insert, the leakage for $\pi^-$ showers drops to about $1\%$, suggesting most of the leakage at $\eta^*$=3.7 is in the transverse direction with a minimal amount of it leaking through the beampipe. This indicates that neutrons, which are produced over a large solid angle, traverse the hole and are measured rather than being lost in the small solid angle of the beampipe exit (as illustrated in Fig.~\ref{fig:showershape}).

\subsection{Single-hadron energy resolution}
\begin{figure}[h]
    \centering
    \includegraphics[width=\columnwidth]{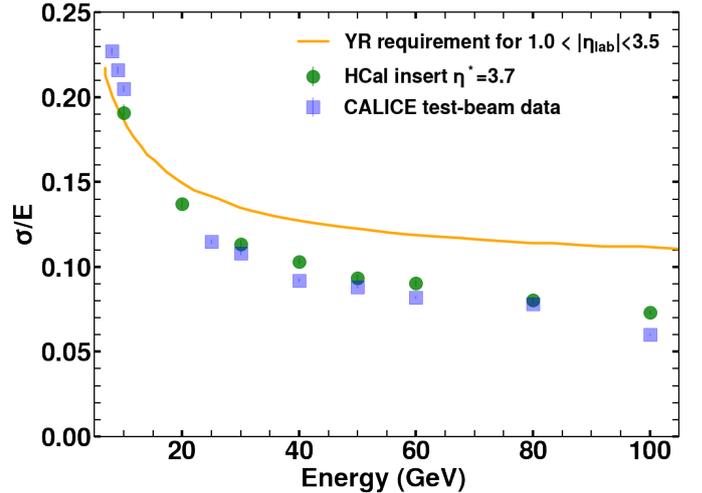}
      \caption{Energy resolution of HCal insert (solid circles, blue online) and the CALICE W-AHCAL test-beam data~\cite{CALICE:2015fpv} (solid squares, blue online) for $\pi^-$ at $\eta^*$=3.7. The EIC Yellow-Report~\cite{AbdulKhalek:2021gbh} requirement is shown by the solid orange line.  }
    \label{res_comp}
\end{figure}

Figure~\ref{res_comp} shows that the resolution, defined as ratio of the width ($\sigma$) to the mean value of the fitted energy distribution, for charged pions at $\eta^*$=3.7. The HCal insert meets the requirements set in the EIC Yellow~Report~\cite{AbdulKhalek:2021gbh}, even beyond the pseudorapidity range for which it was set. 

The expected resolution for the HG-CALI is similar to the CALICE test-beam data~\cite{CALICE:2015fpv}. We have also implemented the CALICE geometry in \textsc{DD4HEP} for both the steel and tungsten prototypes and were able to reproduce the linearity and the resolutions reported in Ref.~\cite{CALICE:2015fpv} to within 10$\%$, indicating that the similarity that we see in Fig.~\ref{res_comp} is not coincidental.

\subsection{Pseudorapidity dependence of performance}
\label{sec:etaperformance}

\begin{figure*}
    \centering
     \includegraphics[width=1\textwidth]{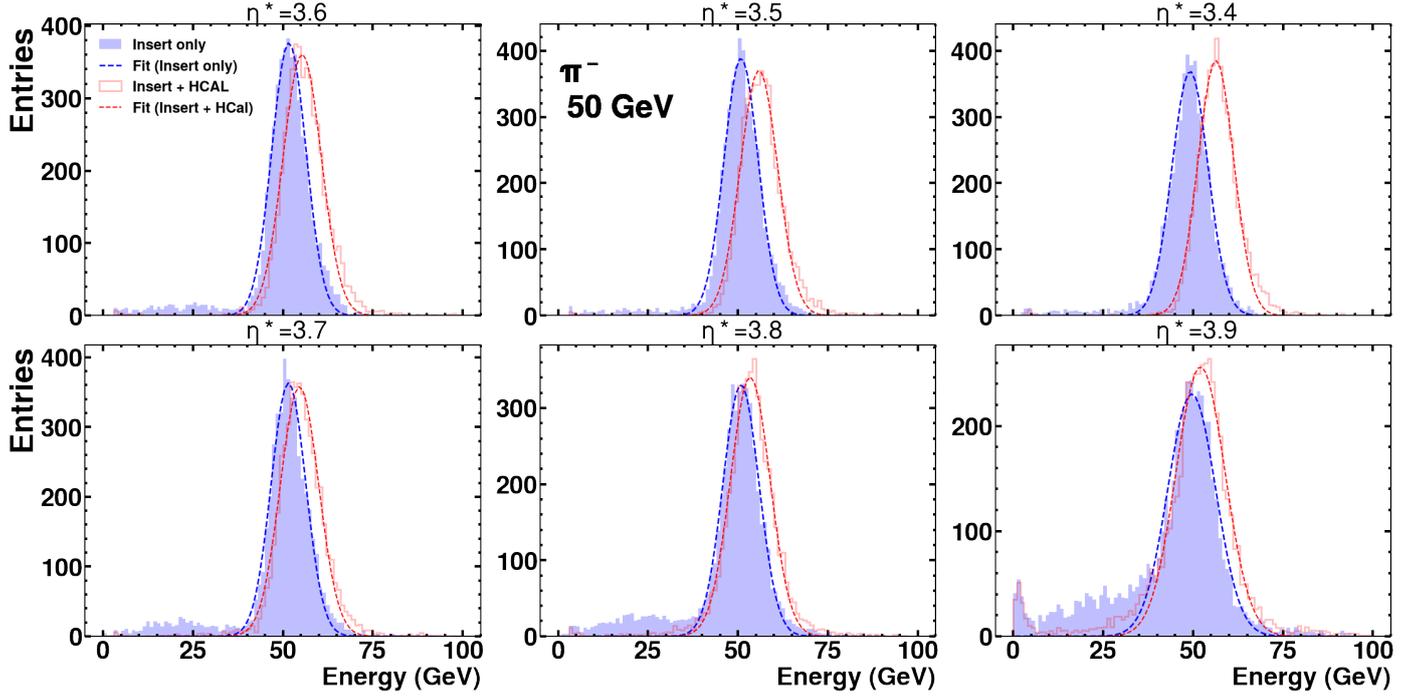}
        \caption{Sum of the hit energies in the calorimeter for $\pi^-$ generated at constant energy 50 GeV but at different pseudorapidity ($\eta^*$). The filled histogram (blue online) is the sum of energy of hits from the HCal insert while the outlined histogram (red online) represent the total energy deposited in the HCal insert and the HCal, weighted with their respective calorimeter sampling fractions. The dashed curves represent Gaussian fits to the energy responses.}
    \label{fig:response_eta}
\end{figure*}

Figure~\ref{fig:response_eta} shows the distribution of shower energies at a constant pion energy of 50 GeV but with different $\eta^*$. The measured distributions using both the HCal insert and HCal can be well described by a Gaussian fit, but the fit quality deteriorates with increasing $\eta^*$. For $\eta^*>$3.9, the fraction of events in the low-energy tail increases due to losses to the beampipe hole. This stems from the worse geometrical acceptance of the HCal insert here (see Sec.~\ref{geometric_acceptance}). 

\begin{figure}
    \centering
    \includegraphics[width=\columnwidth]{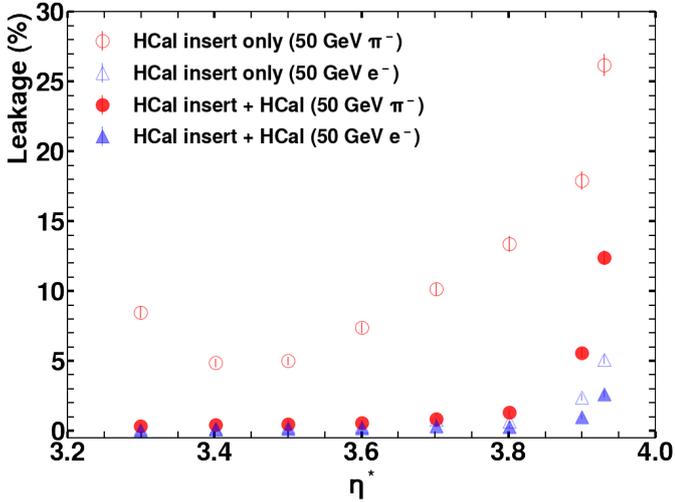}
      \caption{Shower energy leakage as a function of $\eta^{*}$ for 50~GeV pions and electrons. Pions and electrons are shown with red circles and blue triangles, respectively (colored online). The open markers show the leakage when considering only the HCal insert, while the solid markers show the leakage when considering both the HCal and the insert.}
    \label{leakage_eta}
    
\end{figure}

Figure~\ref{leakage_eta} shows the leakage for 50~GeV $\pi^-$ and $e^-$ at different $\eta^{*}$. Taking both the HCal endcap and the HCal insert into account removes most of the leakage, which is in the transverse direction and can be recovered, as shown by solid markers. At the high $\eta^*$ the leakage increases sharply but is still only about 5$\%$ at $\eta^{*}=3.9$ for $\pi^-$ and about 2$\%$ for $e^-$.

\begin{figure}
    \centering
    \includegraphics[width=\columnwidth]{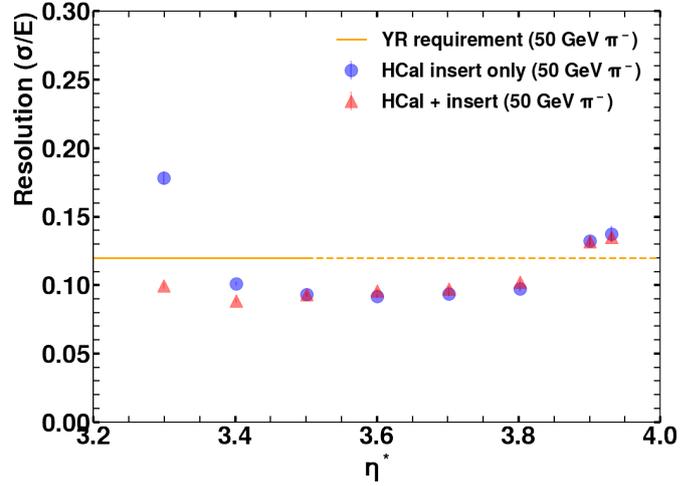}
  \caption{Resolution as a function of pseudorapidity for 50~GeV pions. Resolution is shown when considering only the HCal insert (blue circles) and when considering both the HCal and the insert (red triangles). The YR requirement reported up to $\eta^{*}<$3.5 is shown in solid orange and the extrapolation beyond $\eta^{*}=$3.5  is presented with the dashed line.}
    \label{resolution_eta}
    
\end{figure}

Figure~\ref{resolution_eta} shows the energy resolutions obtained with the HCal insert and the HCal as a function of $\eta^*$ for 50~GeV $\pi^-$. The resolution obtained with HG-CALI only is constant at about $9\%$ over the region $3.4<\eta^*<3.8$ and it degrades at lower and higher $\eta^*$.  Including the HCal improves the resolution only at low $\eta^*$ but not at high $\eta^*$. The horizontal line represents the EIC Yellow Report requirement~\cite{AbdulKhalek:2021gbh}, which is defined up to $\eta=3.5$, beyond which only ``degraded performance'' is specified. The HG-CALI resolution meets the YR requirement up to $\eta^*=3.8$, and gets slightly worse at the highest $\eta^*$ region due to the poorer acceptance as discussed in Sec.~\ref{geometric_acceptance}.

%% file: Conclusions.tex
\section{Summary and conclusions} 
\label{sec:Conclusions}

We have presented a design for a high-granularity calorimeter insert (HG-CALI) for the forward hadronic calorimeter of the EIC detector. Its sampling design is based on the CALICE ``SiPM-on-tile'' concept and uses scintillator megatiles that are grooved to define hexagonal cells. Each layer is distinctly shaped to adapt to the conical beampipe shape that accommodates a beam-crossing angle of 25 mrad. 

The HG-CALI represents the first concrete proposal to instrument the $3<\eta^{*}<4$ range, which is challenging due to the EIC crossing angle. It offers an optimal acceptance coverage that is limited only by the clearance from the beampipe. It thus represents a way to maximize acceptance up to nearly $\eta^{*}=4.0$, which is a key requirement set in the EIC Yellow Report~\cite{AbdulKhalek:2021gbh}.

The HG-CALI is designed to integrate well with the rest of the endcap calorimeters and will enable easy access to the scintillator layers and SiPMs for maintenance and upgrades. Such a feature will enable frequent annealing of the SiPMs, which will keep radiation damage within manageable levels. 

The energy resolution of the HG-CALI meets the EIC Yellow Report requirements even with a basic reconstruction algorithm. Moreover, this device will provide 5D shower information (energy, time, and 3D position) that can be exploited with machine-learning techniques. 

The HG-CALI will complement the forward HCal in the region $3.0<\eta^{*}<3.4$ or so by providing an accurate estimate of the shower start position and incident angle over some range in azimuth. Such information could be used with machine-learning approaches to account for shower fluctuations and thus improve the energy resolution of the combined system.

This design will be complemented by a high-granularity electromagnetic calorimeter insert based on tungsten-powder/scintillating fiber with SiPM readout. The complete insert system will be compensated, yielding the prospect of precise and accurate measurements.

The HG-CALI has the potential to unleash the promise of imaging calorimetry at the EIC, which could support a rich physics program centered on forward jets and their substructure in polarized electron-proton and electron-nucleus collisions.

%% file: main.bbl
\begin{thebibliography}{10}
\expandafter\ifx\csname url\endcsname\relax
  \def\url#1{\texttt{#1}}\fi
\expandafter\ifx\csname urlprefix\endcsname\relax\def\urlprefix{URL }\fi
\expandafter\ifx\csname href\endcsname\relax
  \def\href#1#2{#2} \def\path#1{#1}\fi

\bibitem{AbdulKhalek:2021gbh}
R.~Abdul~Khalek, et~al., {Science Requirements and Detector Concepts for the
  Electron-Ion Collider: EIC Yellow Report} (3 2021).
\newblock \href {http://arxiv.org/abs/2103.05419} {\path{arXiv:2103.05419}}.

\bibitem{Accardi:2012qut}
A.~Accardi, et~al., {Electron Ion Collider: The Next QCD Frontier}:
  {Understanding the glue that binds us all}, Eur. Phys. J. A 52~(9) (2016)
  268.
\newblock \href {http://arxiv.org/abs/1212.1701} {\path{arXiv:1212.1701}},
  \href {https://doi.org/10.1140/epja/i2016-16268-9}
  {\path{doi:10.1140/epja/i2016-16268-9}}.

\bibitem{ATHENA}
{ATHENA Collaboration}, \href{https://doi.org/10.5281/zenodo.6539707}{{ATHENA
  Detector Proposal - A Totally Hermetic Electron Nucleus Apparatus proposed
  for IP6 at the Electron-Ion Collider}} (Dec. 2021).
\newblock \href {https://doi.org/10.5281/zenodo.6539707}
  {\path{doi:10.5281/zenodo.6539707}}.
\newline\urlprefix\url{https://doi.org/10.5281/zenodo.6539707}

\bibitem{ecce_consortium_2022_6537588}
{ECCE consortium}, {EIC Comprehensive Chromodynamics Experiment Collaboration
  Detector Proposal} (May 2022).
\newblock \href {https://doi.org/10.5281/zenodo.6537588}
  {\path{doi:10.5281/zenodo.6537588}}.

\bibitem{Adkins:2022jfp}
J.~K. Adkins, et~al., {Design of the ECCE Detector for the Electron Ion
  Collider} (9 2022).
\newblock \href {http://arxiv.org/abs/2209.02580} {\path{arXiv:2209.02580}}.

\bibitem{Arrington:2021yeb}
J.~Arrington, et~al., {EIC Physics from An All-Silicon Tracking Detector} (2
  2021).
\newblock \href {http://arxiv.org/abs/2102.08337} {\path{arXiv:2102.08337}}.

\bibitem{AbdulKhalek:2022hcn}
R.~Abdul~Khalek, et~al., {Snowmass 2021 White Paper: Electron Ion Collider for
  High Energy Physics}, in: {2022 Snowmass Summer Study}, 2022.
\newblock \href {http://arxiv.org/abs/2203.13199} {\path{arXiv:2203.13199}}.

\bibitem{Sefkow:2015hna}
F.~Sefkow, A.~White, K.~Kawagoe, R.~P\"oschl, J.~Repond, {Experimental Tests of
  Particle Flow Calorimetry}, Rev. Mod. Phys. 88 (2016) 015003.
\newblock \href {http://arxiv.org/abs/1507.05893} {\path{arXiv:1507.05893}},
  \href {https://doi.org/10.1103/RevModPhys.88.015003}
  {\path{doi:10.1103/RevModPhys.88.015003}}.

\bibitem{Thomson:2009rp}
M.~A. Thomson, {Particle Flow Calorimetry and the PandoraPFA Algorithm}, Nucl.
  Instrum. Meth. A 611 (2009) 25--40.
\newblock \href {http://arxiv.org/abs/0907.3577} {\path{arXiv:0907.3577}},
  \href {https://doi.org/10.1016/j.nima.2009.09.009}
  {\path{doi:10.1016/j.nima.2009.09.009}}.

\bibitem{CERN-LHCC-2017-023}
{The Phase-2 Upgrade of the CMS Endcap Calorimeter}, Tech. rep., CERN, Geneva
  (Nov 2017).
\newblock \href {https://doi.org/10.17181/CERN.IV8M.1JY2}
  {\path{doi:10.17181/CERN.IV8M.1JY2}}.

\bibitem{ILDConceptGroup:2020sfq}
H.~Abramowicz, et~al., {International Large Detector: Interim Design Report} (3
  2020).
\newblock \href {http://arxiv.org/abs/2003.01116} {\path{arXiv:2003.01116}}.

\bibitem{Dannheim:2019rcr}
A.~C. Abusleme~Hoffman, et~al., {Detector Technologies for CLIC} 1/2019 (5
  2019).
\newblock \href {http://arxiv.org/abs/1905.02520} {\path{arXiv:1905.02520}},
  \href {https://doi.org/10.23731/CYRM-2019-001}
  {\path{doi:10.23731/CYRM-2019-001}}.

\bibitem{CEPCStudyGroup:2018ghi}
M.~Dong, et~al., {CEPC Conceptual Design Report: Volume 2 - Physics \&
  Detector} (11 2018).
\newblock \href {http://arxiv.org/abs/1811.10545} {\path{arXiv:1811.10545}}.

\bibitem{Li:2021gla}
L.~Li, Q.~Shan, W.~Jia, B.~Yu, Y.~Liu, Y.~Shi, T.~Hu, J.~Jiang, {Optimization
  of the CEPC-AHCAL scintillator detector cells}, JINST 16~(03) (2021) P03001.
\newblock \href {https://doi.org/10.1088/1748-0221/16/03/P03001}
  {\path{doi:10.1088/1748-0221/16/03/P03001}}.

\bibitem{Duan:2021mvk}
Y.~Duan, et~al., {Scintillator tile batch test of CEPC AHCAL}, JINST 17~(05)
  (2022) P05006.
\newblock \href {http://arxiv.org/abs/2111.03660} {\path{arXiv:2111.03660}},
  \href {https://doi.org/10.1088/1748-0221/17/05/P05006}
  {\path{doi:10.1088/1748-0221/17/05/P05006}}.

\bibitem{Simon:2018xzl}
F.~Simon, {Silicon photomultipliers in particle and nuclear physics}, Nucl.
  Instrum. Meth. A 926 (2019) 85--100.
\newblock \href {http://arxiv.org/abs/1811.03877} {\path{arXiv:1811.03877}},
  \href {https://doi.org/10.1016/j.nima.2018.11.042}
  {\path{doi:10.1016/j.nima.2018.11.042}}.

\bibitem{Blazey:2009zz}
G.~Blazey, et~al., {Directly Coupled Tiles as Elements of a Scintillator
  Calorimeter with MPPC Readout}, Nucl. Instrum. Meth. A 605 (2009) 277--281.
\newblock \href {https://doi.org/10.1016/j.nima.2009.03.253}
  {\path{doi:10.1016/j.nima.2009.03.253}}.

\bibitem{Simon:2010hf}
F.~Simon, C.~Soldner, {Uniformity Studies of Scintillator Tiles directly
  coupled to SiPMs for Imaging Calorimetry}, Nucl. Instrum. Meth. A 620 (2010)
  196--201.
\newblock \href {http://arxiv.org/abs/1001.4665} {\path{arXiv:1001.4665}},
  \href {https://doi.org/10.1016/j.nima.2010.03.142}
  {\path{doi:10.1016/j.nima.2010.03.142}}.

\bibitem{Andreev:2004uy}
V.~Andreev, et~al., {A high granularity scintillator hadronic-calorimeter with
  SiPM readout for a linear collider detector}, Nucl. Instrum. Meth. A 540
  (2005) 368--380.
\newblock \href {https://doi.org/10.1016/j.nima.2004.12.002}
  {\path{doi:10.1016/j.nima.2004.12.002}}.

\bibitem{Andreev:2005cua}
V.~Andreev, et~al., {A high-granularity plastic scintillator tile hadronic
  calorimeter with APD readout for a linear collider detector}, Nucl. Instrum.
  Meth. A 564 (2006) 144--154.
\newblock \href {https://doi.org/10.1016/j.nima.2006.04.044}
  {\path{doi:10.1016/j.nima.2006.04.044}}.

\bibitem{CALICE:2010fpb}
C.~Adloff, et~al., {Construction and Commissioning of the CALICE Analog Hadron
  Calorimeter Prototype}, JINST 5 (2010) P05004.
\newblock \href {http://arxiv.org/abs/1003.2662} {\path{arXiv:1003.2662}},
  \href {https://doi.org/10.1088/1748-0221/5/05/P05004}
  {\path{doi:10.1088/1748-0221/5/05/P05004}}.

\bibitem{Simon_2010}
F.~Simon, C.~Soldner, Uniformity studies of scintillator tiles directly coupled
  to {SiPMs} for imaging calorimetry, Nuclear Instruments and Methods in
  Physics Research Section A: Accelerators, Spectrometers, Detectors and
  Associated Equipment 620~(2-3) (2010) 196--201.
\newblock \href {https://doi.org/10.1016/j.nima.2010.03.142}
  {\path{doi:10.1016/j.nima.2010.03.142}}.

\bibitem{Feickert:2021ajf}
M.~Feickert, B.~Nachman, {A Living Review of Machine Learning for Particle
  Physics} (2 2021).
\newblock \href {http://arxiv.org/abs/2102.02770} {\path{arXiv:2102.02770}}.

\bibitem{Paganini:2017dwg}
M.~Paganini, L.~de~Oliveira, B.~Nachman, {CaloGAN : Simulating 3D high energy
  particle showers in multilayer electromagnetic calorimeters with generative
  adversarial networks}, Phys. Rev. D 97~(1) (2018) 014021.
\newblock \href {http://arxiv.org/abs/1712.10321} {\path{arXiv:1712.10321}},
  \href {https://doi.org/10.1103/PhysRevD.97.014021}
  {\path{doi:10.1103/PhysRevD.97.014021}}.

\bibitem{Paganini:2017hrr}
M.~Paganini, L.~de~Oliveira, B.~Nachman, {Accelerating Science with Generative
  Adversarial Networks: An Application to 3D Particle Showers in Multilayer
  Calorimeters}, Phys. Rev. Lett. 120~(4) (2018) 042003.
\newblock \href {http://arxiv.org/abs/1705.02355} {\path{arXiv:1705.02355}},
  \href {https://doi.org/10.1103/PhysRevLett.120.042003}
  {\path{doi:10.1103/PhysRevLett.120.042003}}.

\bibitem{Belayneh:2019vyx}
D.~Belayneh, et~al., {Calorimetry with deep learning: particle simulation and
  reconstruction for collider physics}, Eur. Phys. J. C 80~(7) (2020) 688.
\newblock \href {http://arxiv.org/abs/1912.06794} {\path{arXiv:1912.06794}},
  \href {https://doi.org/10.1140/epjc/s10052-020-8251-9}
  {\path{doi:10.1140/epjc/s10052-020-8251-9}}.

\bibitem{Qasim:2019otl}
S.~R. Qasim, J.~Kieseler, Y.~Iiyama, M.~Pierini, {Learning representations of
  irregular particle-detector geometry with distance-weighted graph networks},
  Eur. Phys. J. C 79~(7) (2019) 608.
\newblock \href {http://arxiv.org/abs/1902.07987} {\path{arXiv:1902.07987}},
  \href {https://doi.org/10.1140/epjc/s10052-019-7113-9}
  {\path{doi:10.1140/epjc/s10052-019-7113-9}}.

\bibitem{DiBello:2020bas}
F.~A. Di~Bello, S.~Ganguly, E.~Gross, M.~Kado, M.~Pitt, L.~Santi, J.~Shlomi,
  {Towards a Computer Vision Particle Flow}, Eur. Phys. J. C 81~(2) (2021) 107.
\newblock \href {http://arxiv.org/abs/2003.08863} {\path{arXiv:2003.08863}},
  \href {https://doi.org/10.1140/epjc/s10052-021-08897-0}
  {\path{doi:10.1140/epjc/s10052-021-08897-0}}.

\bibitem{Buhmann:2020pmy}
E.~Buhmann, S.~Diefenbacher, E.~Eren, F.~Gaede, G.~Kasieczka, A.~Korol,
  K.~Kr\"uger, {Getting High: High Fidelity Simulation of High Granularity
  Calorimeters with High Speed}, Comput. Softw. Big Sci. 5~(1) (2021) 13.
\newblock \href {http://arxiv.org/abs/2005.05334} {\path{arXiv:2005.05334}},
  \href {https://doi.org/10.1007/s41781-021-00056-0}
  {\path{doi:10.1007/s41781-021-00056-0}}.

\bibitem{Akchurin:2021afn}
N.~Akchurin, C.~Cowden, J.~Damgov, A.~Hussain, S.~Kunori, {On the use of neural
  networks for energy reconstruction in high-granularity calorimeters}, JINST
  16~(12) (2021) P12036.
\newblock \href {http://arxiv.org/abs/2107.10207} {\path{arXiv:2107.10207}},
  \href {https://doi.org/10.1088/1748-0221/16/12/P12036}
  {\path{doi:10.1088/1748-0221/16/12/P12036}}.

\bibitem{Pata:2021oez}
J.~Pata, J.~Duarte, J.-R. Vlimant, M.~Pierini, M.~Spiropulu, {MLPF: Efficient
  machine-learned particle-flow reconstruction using graph neural networks},
  Eur. Phys. J. C 81~(5) (2021) 381.
\newblock \href {http://arxiv.org/abs/2101.08578} {\path{arXiv:2101.08578}},
  \href {https://doi.org/10.1140/epjc/s10052-021-09158-w}
  {\path{doi:10.1140/epjc/s10052-021-09158-w}}.

\bibitem{Neubuser:2021uui}
C.~Neub\"user, J.~Kieseler, P.~Lujan, {Optimising longitudinal and lateral
  calorimeter granularity for software compensation in hadronic showers using
  deep neural networks}, Eur. Phys. J. C 82~(1) (2022) 92.
\newblock \href {http://arxiv.org/abs/2101.08150} {\path{arXiv:2101.08150}},
  \href {https://doi.org/10.1140/epjc/s10052-022-10031-7}
  {\path{doi:10.1140/epjc/s10052-022-10031-7}}.

\bibitem{Akchurin:2021ahx}
N.~Akchurin, C.~Cowden, J.~Damgov, A.~Hussain, S.~Kunori, {Perspectives on the
  Calibration of CNN Energy Reconstruction in Highly Granular Calorimeters} (8
  2021).
\newblock \href {http://arxiv.org/abs/2108.10963} {\path{arXiv:2108.10963}}.

\bibitem{Buhmann:2021caf}
E.~Buhmann, S.~Diefenbacher, D.~Hundhausen, G.~Kasieczka, W.~Korcari, E.~Eren,
  F.~Gaede, K.~Kr\"uger, P.~McKeown, L.~Rustige, {Hadrons, better, faster,
  stronger}, Mach. Learn. Sci. Tech. 3~(2) (2022) 025014.
\newblock \href {http://arxiv.org/abs/2112.09709} {\path{arXiv:2112.09709}},
  \href {https://doi.org/10.1088/2632-2153/ac7848}
  {\path{doi:10.1088/2632-2153/ac7848}}.

\bibitem{Khattak:2021ndw}
G.~R. Khattak, S.~Vallecorsa, F.~Carminati, G.~M. Khan, {Fast simulation of a
  high granularity calorimeter by generative adversarial networks}, Eur. Phys.
  J. C 82~(4) (2022) 386.
\newblock \href {http://arxiv.org/abs/2109.07388} {\path{arXiv:2109.07388}},
  \href {https://doi.org/10.1140/epjc/s10052-022-10258-4}
  {\path{doi:10.1140/epjc/s10052-022-10258-4}}.

\bibitem{Chadeeva:2022kay}
M.~Chadeeva, S.~Korpachev, {Machine-learning-based prediction of parameters of
  secondaries in hadronic showers using calorimetric observables} (5 2022).
\newblock \href {http://arxiv.org/abs/2205.12534} {\path{arXiv:2205.12534}}.

\bibitem{Qasim:2022rww}
S.~R. Qasim, N.~Chernyavskaya, J.~Kieseler, K.~Long, O.~Viazlo, M.~Pierini,
  R.~Nawaz, {End-to-end multi-particle reconstruction in high occupancy imaging
  calorimeters with graph neural networks} (4 2022).
\newblock \href {http://arxiv.org/abs/2204.01681} {\path{arXiv:2204.01681}}.

\bibitem{Mikuni:2022xry}
V.~Mikuni, B.~Nachman, {Score-based Generative Models for Calorimeter Shower
  Simulation} (6 2022).
\newblock \href {http://arxiv.org/abs/2206.11898} {\path{arXiv:2206.11898}}.

\bibitem{Repond:2019fbz}
J.~Repond, {Detector Concepts of the Electron-Ion Collider}, PoS High-pT2019
  (2020) 015.
\newblock \href {https://doi.org/10.22323/1.355.0015}
  {\path{doi:10.22323/1.355.0015}}.

\bibitem{core_proto_collaboration_2021_6536630}
{CORE Proto-Collaboration}, \href{https://doi.org/10.5281/zenodo.6536630}{{CORE
  - a COmpact detectoR for the EIC}} (Dec. 2021).
\newblock \href {https://doi.org/10.5281/zenodo.6536630}
  {\path{doi:10.5281/zenodo.6536630}}.
\newline\urlprefix\url{https://doi.org/10.5281/zenodo.6536630}

\bibitem{Bock:2022lwp}
F.~Bock, et~al., {Design and Simulated Performance of Calorimetry Systems for
  the ECCE Detector at the Electron Ion Collider} (7 2022).
\newblock \href {http://arxiv.org/abs/2207.09437} {\path{arXiv:2207.09437}}.

\bibitem{ZEUSCalorimeterGroup:1989ill}
U.~Behrens, et~al., {Test of the {ZEUS} Forward Calorimeter Prototype}, Nucl.
  Instrum. Meth. A 289 (1990) 115--138.
\newblock \href {https://doi.org/10.1016/0168-9002(90)90253-3}
  {\path{doi:10.1016/0168-9002(90)90253-3}}.

\bibitem{H1CalorimeterGroup:1993boq}
B.~Andrieu, et~al., {The H1 liquid argon calorimeter system}, Nucl. Instrum.
  Meth. A 336 (1993) 460--498.
\newblock \href {https://doi.org/10.1016/0168-9002(93)91257-N}
  {\path{doi:10.1016/0168-9002(93)91257-N}}.

\bibitem{CALICE:2010thx}
C.~Adloff, et~al., {Electromagnetic response of a highly granular hadronic
  calorimeter}, JINST 6 (2011) P04003.
\newblock \href {http://arxiv.org/abs/1012.4343} {\path{arXiv:1012.4343}},
  \href {https://doi.org/10.1088/1748-0221/6/04/P04003}
  {\path{doi:10.1088/1748-0221/6/04/P04003}}.

\bibitem{CALICE:2011brp}
C.~Adloff, J.~Blaha, J.~J. Blaising, C.~Drancourt, A.~Espargiliere, R.~Galione,
  N.~Geffroy, Y.~Karyotakis, J.~Prast, G.~Vouters, {Tests of a particle flow
  algorithm with CALICE test beam data}, JINST 6 (2011) P07005.
\newblock \href {http://arxiv.org/abs/1105.3417} {\path{arXiv:1105.3417}},
  \href {https://doi.org/10.1088/1748-0221/6/07/P07005}
  {\path{doi:10.1088/1748-0221/6/07/P07005}}.

\bibitem{CALICE:2012eac}
C.~Adloff, et~al., {Hadronic energy resolution of a highly granular
  scintillator-steel hadron calorimeter using software compensation
  techniques}, JINST 7 (2012) P09017.
\newblock \href {http://arxiv.org/abs/1207.4210} {\path{arXiv:1207.4210}},
  \href {https://doi.org/10.1088/1748-0221/7/09/P09017}
  {\path{doi:10.1088/1748-0221/7/09/P09017}}.

\bibitem{Simon:2013zya}
F.~Simon, C.~Soldner, L.~Weuste, {T3B \textemdash{} an experiment to measure
  the time structure of hadronic showers}, JINST 8 (2013) P12001.
\newblock \href {http://arxiv.org/abs/1309.6143} {\path{arXiv:1309.6143}},
  \href {https://doi.org/10.1088/1748-0221/8/12/P12001}
  {\path{doi:10.1088/1748-0221/8/12/P12001}}.

\bibitem{CALICE:2013osg}
C.~Adloff, et~al., {Track segments in hadronic showers in a highly granular
  scintillator-steel hadron calorimeter}, JINST 8 (2013) P09001.
\newblock \href {http://arxiv.org/abs/1305.7027} {\path{arXiv:1305.7027}},
  \href {https://doi.org/10.1088/1748-0221/8/09/P09001}
  {\path{doi:10.1088/1748-0221/8/09/P09001}}.

\bibitem{CALICE:2013dkh}
C.~Adloff, et~al., {Validation of GEANT4 Monte Carlo Models with a Highly
  Granular Scintillator-Steel Hadron Calorimeter}, JINST 8 (2013) 07005.
\newblock \href {http://arxiv.org/abs/1306.3037} {\path{arXiv:1306.3037}},
  \href {https://doi.org/10.1088/1748-0221/8/07/P07005}
  {\path{doi:10.1088/1748-0221/8/07/P07005}}.

\bibitem{CALICE:2014tgv}
C.~Adloff, et~al., {The Time Structure of Hadronic Showers in highly granular
  Calorimeters with Tungsten and Steel Absorbers}, JINST 9 (2014) P07022.
\newblock \href {http://arxiv.org/abs/1404.6454} {\path{arXiv:1404.6454}},
  \href {https://doi.org/10.1088/1748-0221/9/07/P07022}
  {\path{doi:10.1088/1748-0221/9/07/P07022}}.

\bibitem{CALICE:2014xjq}
B.~Bilki, et~al., {Pion and proton showers in the CALICE scintillator-steel
  analogue hadron calorimeter}, JINST 10~(04) (2015) P04014.
\newblock \href {http://arxiv.org/abs/1412.2653} {\path{arXiv:1412.2653}},
  \href {https://doi.org/10.1088/1748-0221/10/04/P04014}
  {\path{doi:10.1088/1748-0221/10/04/P04014}}.

\bibitem{CALICE:2015fpv}
M.~Chefdeville, et~al., {Shower development of particles with momenta from 15
  GeV to 150 GeV in the CALICE scintillator-tungsten hadronic calorimeter},
  JINST 10~(12) (2015) P12006.
\newblock \href {http://arxiv.org/abs/1509.00617} {\path{arXiv:1509.00617}},
  \href {https://doi.org/10.1088/1748-0221/10/12/P12006}
  {\path{doi:10.1088/1748-0221/10/12/P12006}}.

\bibitem{CALICE:2016nds}
G.~Eigen, et~al., {Hadron shower decomposition in the highly granular CALICE
  analogue hadron calorimeter}, JINST 11~(06) (2016) P06013.
\newblock \href {http://arxiv.org/abs/1602.08578} {\path{arXiv:1602.08578}},
  \href {https://doi.org/10.1088/1748-0221/11/06/P06013}
  {\path{doi:10.1088/1748-0221/11/06/P06013}}.

\bibitem{CALICE:2018ibt}
J.~Repond, et~al., {Hadronic Energy Resolution of a Combined High Granularity
  Scintillator Calorimeter System}, JINST 13~(12) (2018) P12022.
\newblock \href {http://arxiv.org/abs/1809.03909} {\path{arXiv:1809.03909}},
  \href {https://doi.org/10.1088/1748-0221/13/12/P12022}
  {\path{doi:10.1088/1748-0221/13/12/P12022}}.

\bibitem{Sefkow:2018rhp}
F.~Sefkow, F.~Simon, {A highly granular SiPM-on-tile calorimeter prototype}, J.
  Phys. Conf. Ser. 1162~(1) (2019) 012012.
\newblock \href {http://arxiv.org/abs/1808.09281} {\path{arXiv:1808.09281}},
  \href {https://doi.org/10.1088/1742-6596/1162/1/012012}
  {\path{doi:10.1088/1742-6596/1162/1/012012}}.

\bibitem{Tsai_2012}
O.~D. Tsai, et~al., Results of {R}\&{D} on a new construction technique for
  {W}/{ScFi} calorimeters, J. Phys. Conf. Ser. 404 (2012) 012023.
\newblock \href {https://doi.org/10.1088/1742-6596/404/1/012023}
  {\path{doi:10.1088/1742-6596/404/1/012023}}.

\bibitem{Tsai:2015bna}
O.~D. Tsai, et~al., {Development of a forward calorimeter system for the STAR
  experiment}, J. Phys. Conf. Ser. 587~(1) (2015) 012053.
\newblock \href {https://doi.org/10.1088/1742-6596/587/1/012053}
  {\path{doi:10.1088/1742-6596/587/1/012053}}.

\bibitem{Aschenauer:2016our}
E.-C. Aschenauer, et~al., {The RHIC Cold QCD Plan for 2017 to 2023: A Portal to
  the EIC} (2 2016).
\newblock \href {http://arxiv.org/abs/1602.03922} {\path{arXiv:1602.03922}}.

\bibitem{beampipe}
{EIC Detector Managerie: Step file for the detector chamber},
  \url{https://physdiv.jlab.org/EIC/Menagerie/CAD/Beam\%20Pipe/Detector\%20chamber\%20210222.stp}.

\bibitem{eicug_2022_6422182}
EICUG, \href{https://doi.org/10.5281/zenodo.6422182}{Maximizing the scientific
  output of the eic} (Apr. 2022).
\newblock \href {https://doi.org/10.5281/zenodo.6422182}
  {\path{doi:10.5281/zenodo.6422182}}.
\newline\urlprefix\url{https://doi.org/10.5281/zenodo.6422182}

\bibitem{CORE:2022rso}
R.~Alarcon, et~al., {CORE -- a COmpact detectoR for the EIC} (9 2022).
\newblock \href {http://arxiv.org/abs/2209.00496} {\path{arXiv:2209.00496}}.

\bibitem{Anderle:2021wcy}
D.~P. Anderle, et~al., {Electron-ion collider in China}, Front. Phys. (Beijing)
  16~(6) (2021) 64701.
\newblock \href {http://arxiv.org/abs/2102.09222} {\path{arXiv:2102.09222}},
  \href {https://doi.org/10.1007/s11467-021-1062-0}
  {\path{doi:10.1007/s11467-021-1062-0}}.

\bibitem{STAR:2002ymp}
M.~Beddo, et~al., {The STAR barrel electromagnetic calorimeter}, Nucl. Instrum.
  Meth. A 499 (2003) 725--739.
\newblock \href {https://doi.org/10.1016/S0168-9002(02)01970-8}
  {\path{doi:10.1016/S0168-9002(02)01970-8}}.

\bibitem{CERN-LHCC-97-031}
{The CMS hadron calorimeter project: Technical Design Report}, Technical design
  report. CMS, CERN, Geneva, 1997.

\bibitem{Belloni:2021kcw}
A.~Belloni, et~al., {Test beam study of SiPM-on-tile configurations}, JINST
  16~(07) (2021) P07022.
\newblock \href {http://arxiv.org/abs/2102.08499} {\path{arXiv:2102.08499}},
  \href {https://doi.org/10.1088/1748-0221/16/07/P07022}
  {\path{doi:10.1088/1748-0221/16/07/P07022}}.

\bibitem{deSilva:2020mak}
L.~M.~S. de~Silva, F.~Simon, {Effects of misalignment on response uniformity of
  SiPM-on-tile technology for highly granular calorimeters}, JINST 15~(06)
  (2020) P06030.
\newblock \href {http://arxiv.org/abs/2004.05066} {\path{arXiv:2004.05066}},
  \href {https://doi.org/10.1088/1748-0221/15/06/P06030}
  {\path{doi:10.1088/1748-0221/15/06/P06030}}.

\bibitem{Jiang:2020rhv}
J.~Jiang, S.~Zhao, Y.~Niu, Y.~Shi, Y.~Liu, D.~Han, T.~Hu, B.~Yu, {Study of SiPM
  for CEPC-AHCAL}, Nucl. Instrum. Meth. A 980 (2020) 164481.
\newblock \href {https://doi.org/10.1016/j.nima.2020.164481}
  {\path{doi:10.1016/j.nima.2020.164481}}.

\bibitem{Garutti:2018hfu}
E.~Garutti, Y.~Musienko, {Radiation damage of SiPMs}, Nucl. Instrum. Meth. A
  926 (2019) 69--84.
\newblock \href {http://arxiv.org/abs/1809.06361} {\path{arXiv:1809.06361}},
  \href {https://doi.org/10.1016/j.nima.2018.10.191}
  {\path{doi:10.1016/j.nima.2018.10.191}}.

\bibitem{crossangle_1}
J.~Adam, et~al., \href{https://doi.org/10.5281/zenodo.6514605}{{Accelerator and
  beam conditions critical for physics and detector simulations for the
  Electron-Ion Collider}} (2021).
\newblock \href {https://doi.org/10.5281/zenodo.6514605}
  {\path{doi:10.5281/zenodo.6514605}}.
\newline\urlprefix\url{https://doi.org/10.5281/zenodo.6514605}

\bibitem{crossangle_2}
B.~Schmookler, et~al., \href{https://arxiv.org/abs/2207.04378}{{High $Q^2$
  electron-proton elastic scattering at the future Electron-Ion Collider}}
  (2022).
\newblock \href {https://doi.org/10.48550/ARXIV.2207.04378}
  {\path{doi:10.48550/ARXIV.2207.04378}}.
\newline\urlprefix\url{https://arxiv.org/abs/2207.04378}

\bibitem{Frank:2014zya}
M.~Frank, F.~Gaede, C.~Grefe, P.~Mato, {DD4hep: A Detector Description Toolkit
  for High Energy Physics Experiments}, J. Phys. Conf. Ser. 513 (2014) 022010.
\newblock \href {https://doi.org/10.1088/1742-6596/513/2/022010}
  {\path{doi:10.1088/1742-6596/513/2/022010}}.

\bibitem{GEANT4:2002zbu}
S.~Agostinelli, et~al., {GEANT4--a simulation toolkit}, Nucl. Instrum. Meth. A
  506 (2003) 250--303.
\newblock \href {https://doi.org/10.1016/S0168-9002(03)01368-8}
  {\path{doi:10.1016/S0168-9002(03)01368-8}}.

\bibitem{Birks:1951boa}
J.~B. Birks, {Scintillations from Organic Crystals: Specific Fluorescence and
  Relative Response to Different Radiations}, Proc. Phys. Soc. A 64 (1951)
  874--877.
\newblock \href {https://doi.org/10.1088/0370-1298/64/10/303}
  {\path{doi:10.1088/0370-1298/64/10/303}}.

\bibitem{Graf:2022lwa}
C.~Graf, F.~Simon, {Time-assisted energy reconstruction in a highly-granular
  hadronic calorimeter} (3 2022).
\newblock \href {http://arxiv.org/abs/2203.01317} {\path{arXiv:2203.01317}}.

\bibitem{Qasim:2021hex}
S.~R. Qasim, K.~Long, J.~Kieseler, M.~Pierini, R.~Nawaz, {Multi-particle
  reconstruction in the High Granularity Calorimeter using object condensation
  and graph neural networks}, EPJ Web Conf. 251 (2021) 03072.
\newblock \href {http://arxiv.org/abs/2106.01832} {\path{arXiv:2106.01832}},
  \href {https://doi.org/10.1051/epjconf/202125103072}
  {\path{doi:10.1051/epjconf/202125103072}}.

\bibitem{sketchupModel}
S.~J. Paul, M.~Arratia, O.~Tsai,
  \href{https://doi.org/10.5281/zenodo.6950495}{{Model for a Calorimeter Insert
  for the Electron-Ion Collider}} (7 2022).
\newblock \href {https://doi.org/10.5281/zenodo.6950495}
  {\path{doi:10.5281/zenodo.6950495}}.
\newline\urlprefix\url{https://doi.org/10.5281/zenodo.6950495}

\bibitem{Milton_EIC_pEndcap_Insert_2022}
R.~Milton, B.~Schmookler, Z.~Ji, B.~Karki, S.~Paul, M.~Arratia,
  \href{https://doi.org/10.5281/zenodo.6836014}{{EIC pEndcap Insert in DD4hep}}
  (7 2022).
\newblock \href {https://doi.org/10.5281/zenodo.6836014}
  {\path{doi:10.5281/zenodo.6836014}}.
\newline\urlprefix\url{https://doi.org/10.5281/zenodo.6836014}

\end{thebibliography}
